\documentclass[12pt,a4paper]{article}
\usepackage{jcappub}
\usepackage{newtxtext,newtxmath}
\usepackage{color,xcolor}
\usepackage{graphicx}	
\usepackage{latexsym,amsmath,amssymb}
\usepackage{multicol}
\usepackage{subfigure,float}
\usepackage{dblfloatfix}

\DeclareRobustCommand{\VAN}[3]{#2}
\let\VANthebibliography\thebibliography
\def\thebibliography{\DeclareRobustCommand{\VAN}[3]{##3}\VANthebibliography}

\usepackage{array}
\newcolumntype{P}[1]{>{\centering\arraybackslash}p{#1}}

\def \xhi {\bar{x}_{\rm HI}}
\def \xhic {\bar{x}_{{\rm HI}_{\rm crit}}}
\def \rhi {\rho_{\rm HI}}
\def \nc {$N_{\rm C}$}
\def \HI{{\sc Hi}}
\def \HII{{\sc Hii}}
\def \ff {\rm FF}
\def \mpc {\rm Mpc}

\newcommand{\be}{\begin{equation}}
\newcommand{\e}{\end{equation}}
\newcommand{\bear}{\begin{eqnarray}}
\newcommand{\ear}{\end{eqnarray}}

\newcommand{\comment}[1]{}

\title{Distinguishing reionization models using the largest cluster statistics of the 21-cm maps}

\author[a]{Aadarsh Pathak,}\emailAdd{aadarshbritia@gmail.com}
\author[b]{Satadru Bag,} \emailAdd{satadru@kasi.re.kr}
\author[a]{Saswata Dasgupta,}
\author[a,c]{Suman Majumdar,} \emailAdd{mid.suman@gmail.com}
\author[d]{Rajesh Mondal,}
\author[a]{Mohd Kamran,}
\author[e]{Prakash Sarkar,}

\affiliation[a]{Department of Astronomy, Astrophysics \& Space Engineering, Indian Institute of Technology Indore, Indore 453552, India}

\affiliation[b]{Korea Astronomy and Space Science Institute, Daejeon, Republic of Korea}
\affiliation[c]{Department of Physics, Blackett Laboratory, Imperial College, London SW7 2AZ, U. K.}
\affiliation[d]{Department of Astrophysics, School of Physics and Astronomy, Tel Aviv University, Tel Aviv 69978, Israel}
\affiliation[e]{Ramakrishna Mission English School, Sidhgora, Jamshedpur}

\date{\today}

\abstract{
The evolution of topology and morphology of ionized or neutral hydrogen during different stages of the Epoch of Reionization (EoR) have the potential to provide us a great amount of information about the properties of the ionizing sources during this era. We compare a variety of reionization source models in terms of the geometrical properties of the ionized regions. We show that the {\em percolation transition} in the ionized hydrogen, as studied by tracing the evolution of the Largest Cluster Statistics (LCS), is a robust statistic that can distinguish the fundamentally different scenarios -- inside-out and outside-in reionization. Particularly, the global neutral fraction at the onset of percolation is significantly higher for the inside-out scenario as compared to that for the outside-in reionization. In complementary to percolation analysis, we explore the shape and morphology of the ionized regions as they evolve in different reionization models in terms of the Shapefinders (SFs) that are ratios of the Minkowski functionals (MFs). The shape distribution can readily discern the reionization scenario with extreme non-uniform recombination in the IGM, such as the clumping model. In the rest of the reionization models, the largest ionized region abruptly grows only in terms of its third SF -- `length' -- during percolation while the first two SFs -- `thickness' and `breadth' -- almost remain stable. Thus the ionized hydrogen in these scenarios becomes highly filamentary near percolation and exhibit a `characteristic cross-section' that varies among the source models. Therefore, the geometrical studies based on SFs, together with the percolation analysis can shed light on the reionization sources.
}


\begin{document}
\maketitle

\section{Introduction}

In the history of our Universe, the Epoch of Reionization (EoR) is the period when the neutral hydrogen (\HI) in the Inter-Galactic Medium (IGM) was gradually ionized by the radiations from the first sources \citep[see e.g.][]{furlanetto06,Choudhury09,pritchard12}. In recent times, although the EoR has obtained a boost of attention, still our present understanding of this epoch is very limited due to the scarcity of observations of this epoch. So far, it has been  possible to obtain a limited amount of insight into this epoch via a number of indirect observations such as the CMBR \cite{komatsu11,planck16}, quasar absorption spectra of the Ly$\alpha$ photons at high redshifts \cite{fan03,goto11,becker15,barnett17} and the luminosity function and clustering properties of the Ly$\alpha$ emitters \citep{Ouchi10,ota17,zheng17}.
These observations suggest that the \HI\ reionization was an extended process and had most likely ended by redshift $z \sim 6$. Since these observations can not trace the \HI\ distribution at different IGM ionization stages, thus are unable to resolve a number of fundamental issues such as the topological evolution of the \HI\ distribution in the IGM due to ionization, the morphology of the ionized regions at a particular  reionization stage and how the characteristics of different ionizing sources affect the time evolution of the \HI\ distribution.

The \HI\ 21-cm signal, originating when the electron and proton in the ground state of \HI\ change their spin states from parallel to anti-parallel, promises to  act as a direct probe of the \HI\ distribution in the IGM and thus can potentially answer many of these fundamental issues related to the EoR \citep{furlanetto06,pritchard12}. This signal can directly probe the \HI\ distribution in the IGM at different cosmic times and hence in principle can trace the reionization history. In order to detect this signal, a number of first generation radio telescopes such as GMRT\footnote{\url{http://www.gmrt.ncra.tifr.res.in}} \citep{Paciga13}, LOFAR\footnote{\url{http://www.lofar.org/}} \citep{mertens20}, MWA\footnote{\url{http://www.mwatelescope.org/}} \citep{barry19}, PAPER\footnote{\url{http://eor.berkeley.edu/}} \citep{kolopanis19} and HERA\footnote{\url{https://reionization.org/}} \citep{deboer17,HERA21} are operational. The presence of $\sim 4$--$5$ order of magnitude stronger foreground emission \citep[e.g.][]{dimatteo02,ali08,jelic08,ghosh12} compared to the expected EoR 21-cm signal and various instrumental effects such as system noise \citep{morales05,mcquinn06}, introduce observational obstacles, thereby the interferometric detection of the \HI\ 21-cm signal has not been possible yet. These telescopes are targeting to detect the signal via Fourier statistics instead of trying to make tomographic images of the EoR 21-cm signal. The future Square Kilometre Array (SKA)\footnote{\url{http://www.skatelescope.org/}} \citep{koopmans15, mellema15} is expected to have enough sensitivity to make high-resolution tomographic images of the EoR 21-cm signal. Once the tomographic images are produced through these  future observations, it will open up new avenue for better insights into the EoR, the images will contain both the amplitude and phase information of the 21-cm field.

So far the analysis of EoR 21-cm signal is mostly done via  various Fourier statistics such as the power spectrum \citep{bharadwaj04,barkana05,lidz08,mao12, majumdar13,Majumdar14, majumdar16, pober14,mondal15,mondal16,patil17,giri19b,mertens20,Rahul22}, multi-frequency angular power spectrum \citep{datta07a,mondal18,mondal19,mondal20a}, bispectrum \citep{majumdar18,hutter19,majumdar20,watkinson21,saxena20,kamran21b,kamran21a,mondal21, tiwari21} etc. The EoR 21-cm signal fluctuations, which are mainly determined by the sizes, distributions and connectivity between the ionized regions, is expected to be highly non-Gaussian \citep{bharadwaj05a,mellema06,mondal15}. Thus one would expect the signal statistics, which are of higher order than the power spectrum, to contain more details about this non-Gaussian signal \citep{majumdar18,kamran21b}. However, the estimation of higher order statistics is computationally more involved and their interpretation is also difficult. Further, these higher order Fourier statistics contain information about the fluctuations of the signal at different length scales and correlation between them. However, by definition, they do not contain the phase information of the signal.

There are a number of complementary methods which deal with the signal in real space and have been employed directly on the simulated tomographic images to probe the morphology of the 21-cm field and its evolution during the EoR through the analysis of topology and geometry of this field. Among them, the widely used methods are the Minkowski Functionals (MFs) \citep{Spina21,Friedrich11,hong14,Yoshiura17,Bag18,bag19} which have been used to track the reionization history, the Minkowski tensors \citep{kapahtia19} which are the generalized tensorial form of MFs and can encapsulate the direction information. These are also methods based on percolation theory \citep{Iliev2006,Iliev14,Furlanetto16,Bag18,bag19}, granulometry \citep{Koki17} and persistence theory \citep{elber19} which have been used for the theoretical study of the topological phases of \HII\ regions during EoR. In addition to these, a method based on the Betti numbers \citep{giri20,kapahtia21} provide the number of connected regions, tunnels and cavities to describe the state of the IGM. A recent study of \cite{gorce21} used a method of local variance which probes the reionization history of the observed patches of the sky as well as trace the ionization morphology. However, in most of these earlier works it is widely been considered that modelled reionizing sources are standard fiducial in nature and the inference drawn through these methods depend on the detection of a large number of \HII\ regions at any stage consisting of wide variety of sizes. A recent study of \cite{Bag18} on the contrary focuses on the detection of only the largest ionized region in order to get better insights of the percolation process. For this they have used the ``largest cluster statistic'' along with a shape finding algorithm.  

The insights about the percolation of \HII\ regions, from its onset to the stage when all of the ionized regions are interconnected to form a single large cluster, will depend on the ionizing source and IGM properties. Motivated by this, in this work, we aim to study the morphology and topology of the largest ionized regions (at different stages of the EoR) while considering a number of reionization scenarios with different source and IGM properties. For this purpose, we consider different simulated reionization scenarios. The source models in these simulated reionization scenarios are different from each other in two fundamental ways: a) how the number of ionizing photons emitted by the sources are  related to their host halo mass and b) how the rest frame energy of the ionizing photons are distributed. For all of these scenarios we follow the evolution of the topology of the largest ionized region with the cosmic time. For this, we use the percolation technique in addition with the Shapefinders which are defined as the ratio of the Minkowski functional \citep{Sahni:1998cr}. These Shapefinders can be used to analyze the shape of the ionized regions. Thus, this method allows us to gather information related to individual ionized regions along with the information about how these regions form very large interconnected network at the cosmological scales. For our analysis, we have employed the advanced shape diagnostic tool {\em SURFGEN2}\citep{bag19,Bag:2021hxm}. 

This paper is organized as follows: In section \ref{sec:sim}, we discuss the simulated EoR 21-cm maps for various reionization scenarios which we have used for this work. Section \ref{sec:methods} briefly describes our methods for analysing these simulated 21-cm maps, including the percolation, Minkowski functionals and Shapefinders along with the SURFGEN2 code. In section \ref{sec:results}, we discuss our findings regarding the evolution of largest ionized region. Finally, in section \ref{sec:conclusion}, we summarize our result.

Throughout the paper, we have used the cosmological parameters satisfying the WMAP five year data release $h=0.7$, $\Omega _m =0.27$, $\Omega _{\Lambda} =0.73$, $\Omega _b h^2 =0.0226$ \citep{Komatsu09}.

\section{Simulation}
\label{sec:sim}
Simulating the reionization is essentially a challenging task due to the requirements of high dynamic range in terms of length scale
and mass that one has to take into account. The fundamental problem in simulating reionization from the first principles is essentially to capture the large scale cosmology and small scale astrophysics. This requires one to simulate the reionization in large enough cosmological volume ($\sim 1\, {\rm Gpc}^3$) so that the impact of large scale matter density fluctuations are properly taken into account. At the same time, it requires one to resolve the sources of reionization (typical galaxies; $\sim 10\,{\rm kpc}$ in size) so that their properties are mimicked correctly.

One can use a 3D radiative transfer simulation which works on the principle of ray tracing by following the ionization fronts in the IGM and based on which one can check for different physical processes taking place during the EoR \citep{Ricotti02,Thomas09,Iliev14, Gnedin14,ghara15a}. 
But it is almost computationally impossible to explore the multi-dimensional reionization parameter space using these simulations as they require a humongous amount of computing resources. So, in order to reduce the computational complexities, one can fairly choose the semi-numerical technique \citep{Zahn07, Choudhury09, mesinger11, majumdar13, Majumdar14, mondal15,mondal17} which generally compares the average number of photons with the average number of neutral hydrogen present in a smoothing volume rather than performing a full radiative transfer calculations. These simulations are based on the excursion set formalism proposed by \cite{Furlanetto04} where the ionization map generated at any redshift is mostly dependent on the underlying matter distribution and collapsed structures at that redshift. In this paper, we have used the semi-numerically simulated 21-cm maps from \cite{majumdar16} for our analysis.

The semi-numerical method used in \cite{majumdar16} mainly involves three steps. First, it generates the dark matter distribution at any desired redshift by using $N$-body dark matter gravity only simulations. Second, it identifies the collapsed dark matter halos within the simulated matter distribution which can be accomplished with the help of algorithms like FoF or spherical smoothing. Finally, it considers the halos as the most probable hosts of the sources of ionizing photons and uses excursion set formalism to generate the ionization field which is later converted into the 21-cm field. The authors of \cite{majumdar16} have simulated the signal in a cube with $500\, h^{-1}{\rm Mpc} = 714$ Mpc (in comoving scale) in length along each side. The underlying $N$-body output for their simulation used {\sf CUBEP$^3$M} code \citep{Harnois13} which ran as a part of the PRACE4LOFAR project (PRACE projects 2012061089 and 2014102339). For reionization modeling, they have used $6912^3$ particles of mass $4.0 \times 10^7 M\textsubscript{\(\odot\)}$ on a $13824^3$ mesh (of resolution $0.052$ Mpc) and the simulated matter and halo fields are then interpolated on a $600^3$ grid points (of resolution $1.19$ Mpc). The final 21-cm maps which are used in this analysis also have the same resolution ($1.19$ Mpc). The minimum mass of the halos that can host luminous sources, used in these reionization simulations, is $ 2.09 \times 10^9  M\textsubscript{\(\odot\)}$.

\begin{figure*}
    \includegraphics[width=\linewidth]{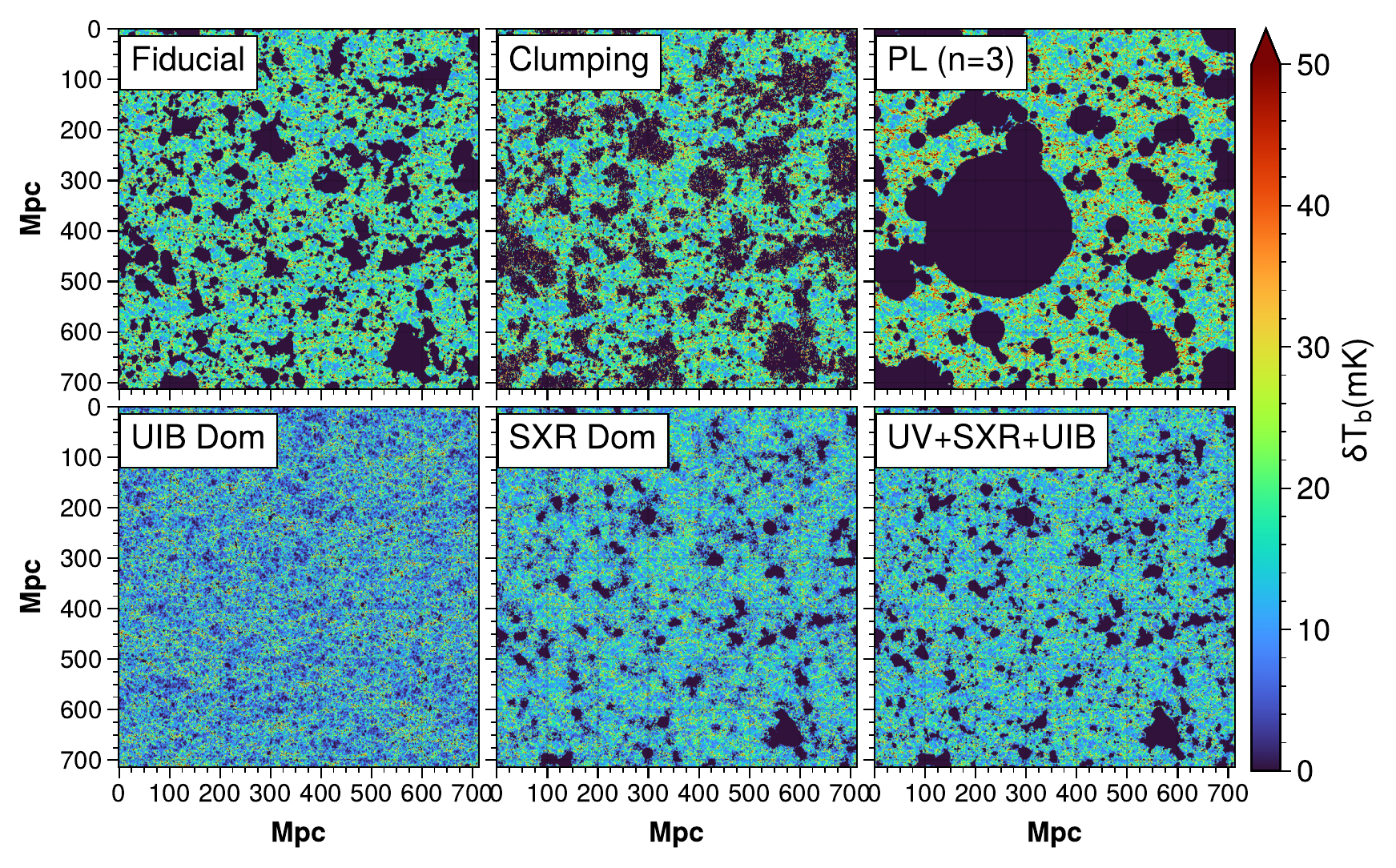}\par 
    \caption{21-cm map of six reionization scenarios taken from \cite{majumdar16} at neutral fraction $\xhi=0.5$.}
    \label{fig:21-cm}
\end{figure*}

\subsection{Source Models and Reionization scenarios}
The evolution of ionized regions in the IGM depends on the characteristics of the sources producing ionizing photons and IGM properties. Thus depending upon the type of possible source characteristics and IGM properties one can construct several reionization scenarios. In this paper, we are using six such simulated reionization scenarios from \cite{majumdar16}. These scenarios are chosen for this analysis for the topologically distinct nature of their resulting 21-cm maps. In the following section, we describe the four sources models which are taken in different combinations to simulate these six scenarios. For a detailed discussion on the same we refer the interested reader to \cite{majumdar16}.

\subsubsection{Source models}
The four source models shaping the six reionization scenarios considered here are:
\begin{enumerate}
    \item \textit{Ultraviolet photons (UV photons):} The galaxies residing in the collapsed dark matter halos are considered to be the most probable sources to produce ionizing photons in the form of UV radiation. Here, it is assumed that the total number of emitted ionizing photons follow the relation :
        \begin{equation}
             N_{\gamma}(M_h) = N_{{\rm ion}} \frac{M_h \Omega_b}{m_p \Omega_m} \label{eq:ng}
        \end{equation}
        where, $N_{{\rm ion}}$ represents the number of photons entering in the IGM per baryon in collapsed objects, $M_h$ is the halo mass and $m_p$ is the mass of proton.
    \item \textit{Uniform Ionizing Background (UIB photons):}  This source model assumes the sources like AGNs, X-ray binaries etc as the most probable ones to produce the hard X-ray photons, following a similar relationship as Equation \eqref{eq:ng}, which due to its long mean free path (comparable to the size of the simulation volume), will contribute to a uniform ionizing photon distribution, where one will not be able to connect a specific photon to the host halo from where it has originated. If this is the only type of sources available in a reionization scenario, it will ultimately lead to a uniform ionizing photon background.\footnote{Note that the effective implementation of this model leads to a uniform ionizing photon distribution, independent of the source locations. Thus the statement that the majority of the photons produced by the sources being hard X-ray with infinite mean free path is an approximate one.}
    \item \textit{Soft X-ray photons (SXR):} This source model produces soft X-ray photons, following a formalism similar to Equation \eqref{eq:ng}, which creates uniform ionizing photon distribution within a region limited by the mean free path of those photons. The mean free path of these soft X-ray photons are determined following the prescription of \citep{mcquinn12}, which is dependent on the redshift of their origin and the frequency of the photon. For simplicity it has been further assumed that all of these soft X-ray photons have same energy ($200$ eV). This implies that photons from a specific source will be uniformly distributed within a spherical region around that source. The radius of this sphere will be determined by the mean free path of the soft X-ray photons.\footnote{Note that a more realistic implementation of the soft X-ray photon mean free path would be a exponentially decreasing ionizing photon field around the source, which extends up to the typical mean free path of the photons.}
    \item \textit{Power Law mass dependent efficiency (PL) :} Here, the number of ionizing UV photons produced by the sources residing in the collapsed dark matter halos is proportional to the $n^{th}$ power of the halo mass following the relation :
    
    \begin{equation}
        N_\gamma(M_h) \propto M^{n}_h \,. 
        \label{eq:PL}
    \end{equation}
     In the scenario considered here, the chosen value of power law index is $3$. Here, $N_\gamma(M_h)$ is the total number of ionizing photons emitted and  $M_h$ is the corresponding halo with mass $M$.
\end{enumerate}

\subsubsection{Reionization Scenarios}
We use the 21-cm maps of six different reioniziation scenarios from \cite{majumdar16} for our analysis here. These scenarios are built by different combinations of the above mentioned source models and their main characteristics are summarized in the table \ref{tab:eor_scenarios}.
\begin{table}
    \centering
   \begin{tabular}{ |P{3.3cm}||p{1.8cm}|P{1.5cm}|P{1.5cm}|P{1.5cm}| }
 \hline
     \textbf{Reionzation scenarios} & \textbf{UV}  & \textbf{UIB} & \textbf{SXR} & \textbf{PL,n}   \\
 \hline  \hline
  Fiducial & 100\% & - & - & 1  \\
 \hline
  Clumping & 100\% \& Non Uniform Recomb. & - & - & 1\\
 \hline
 UIB Dominated & 20\% & 80\% & - & 1  \\
 \hline
 SXR Dominated & 20\% & - & 80\% & 1 \\
 \hline
 UV + SXR + UIB & 50\% & 10\% & 40\% & 1 \\
 \hline
 PL  (n=3) & 100\% & - & - & 3 \\
 \hline
\end{tabular}
\caption{Contribution of different source models in our reionization scenarios. Table taken from \cite{majumdar16}.}
    \label{tab:eor_scenarios}
  
\end{table}

The fiducial scenario considers $100 \%$ ionizing UV photon contribution from the galaxies residing in the halos of mass $\ge 2.09 \times 10^9  M\textsubscript{\(\odot\)}$. In a similar fashion, the clumping and PL ($n=3$) scenarios also consider $100 \%$ UV photon contribution from halos but are slightly different from the fiducial. In all of these scenarios, except one, a density independent uniform rate of recombination have been assumed. Clumping is the only scenario where non-uniform density dependent recombination has been taken into account. In a realistic situation recombination rate will depend on the density of the ionized gas. Dense structures much smaller (few kpc) than the resolution of this simulation are expected to boost the recombination rate significantly. This will lead to the formation of self-shielded regions like
Lyman limit systems. The effect of the presence of these self-shielded regions have been included following the sub-grid prescription of \citep{Choudhury09}. This approach somewhat overestimates the recombination rate but it serves the purpose of generating a topologically distinct EoR 21-cm map. On the other hand, in PL ($n=3$) scenario, high mass halos have a higher weightage in UV photon production than low mass halos as they follow Equation \eqref{eq:PL} instead of \eqref{eq:ng} with $n>1$. Scenarios like UIB dominated, SXR dominated and UV+SXR+UIB have a mixed contribution of different types of ionizing photons as tabulated in table \ref{tab:eor_scenarios}. The UIB dominated and SXR dominated scenarios have $80 \%$ of ionizing photons in the form of hard and soft X-rays respectively. In all of the scenarios considered here ionization state of a resolution element of the simulation is determined by comparing the smoothed ionizing photon field (dependent on the source model and ionizing efficiency of the sources) and the smoothed neutral hydrogen field following the excursion set formalism. The smoothing radius is varied from the resolution element of the simulation up to a predefined upper limit determined generally by the mean free path of the ionizing photons. However, if the ionization condition (i.e. ionizing photon density is greater than the neutral hydrogen density) at a simulation pixel is not satisfied for any such smoothing radii, one assigns an ionization fraction to that cell by simply taking the ratio of the number of ionizing photons in the cell divided by the number of neutral hydrogen atoms in that cell. Number of cells in the simulation volume with fractional ionization becomes particularly prominent in case of the UIB dominated and SXR dominated scenarios.

In this work, while analyzing the topology of the 21-cm maps, we consider the completely ionized regions only\footnote{We consider the regions with $\rho_{\HI}({\bf x})>0$ as `neutral' which includes partially ionized regions too.}, defined as $\rho_{\rm HI}({\bf x})=0$ in the simulation\footnote{In reality, the low frequency radio interferometers will measure the brightness temperature which is proportional to the neutral hydrogen density after mean subtraction, i.e. $(\rhi({\bf x})-\bar{\rhi})$. The ionized region, defined here as $\rho_{\rm HI}({\bf x})=0$, would be mapped to the minima of the observed brightness temperature field that can be easily traced in the high quality data with marginal noise.}. The mass-averaged neutral fraction ($\xhi$) and ionized filling factor (FF) at a specific redshift are defined as 
\begin{equation} 
\xhi(z) =\bar{\rho}_{\rm HI}(z)/\bar{\rho}_{\rm H}(z) \;,
\end{equation}
and
\begin{equation} 
    {\rm FF} = \frac{\rm total ~volume ~of ~all ~the ~ionized ~regions}{{\rm simulation ~volume} } \label{eq:FF}
\end{equation}
respectively. In other words, FF and $(1-\xhi)$ essentially measure the fraction of hydrogen mass and volume that is ionized at a given redshift/time.

\begin{figure}
\includegraphics[scale=0.3
]{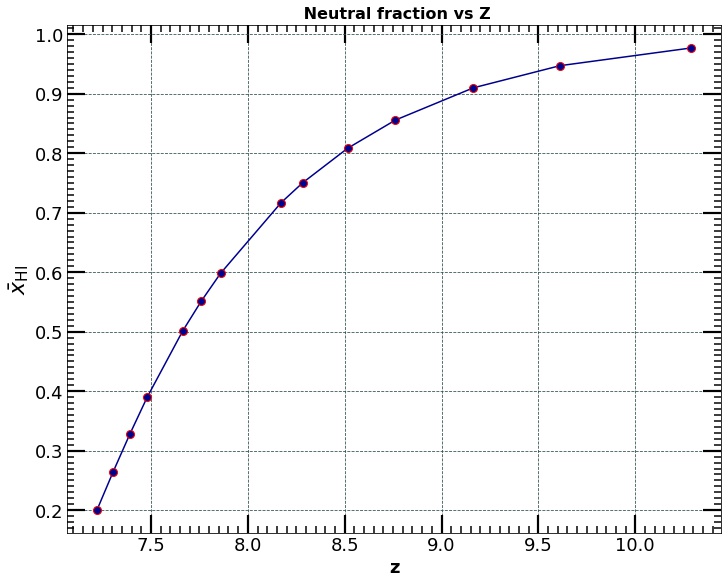}
\includegraphics[scale=0.6]{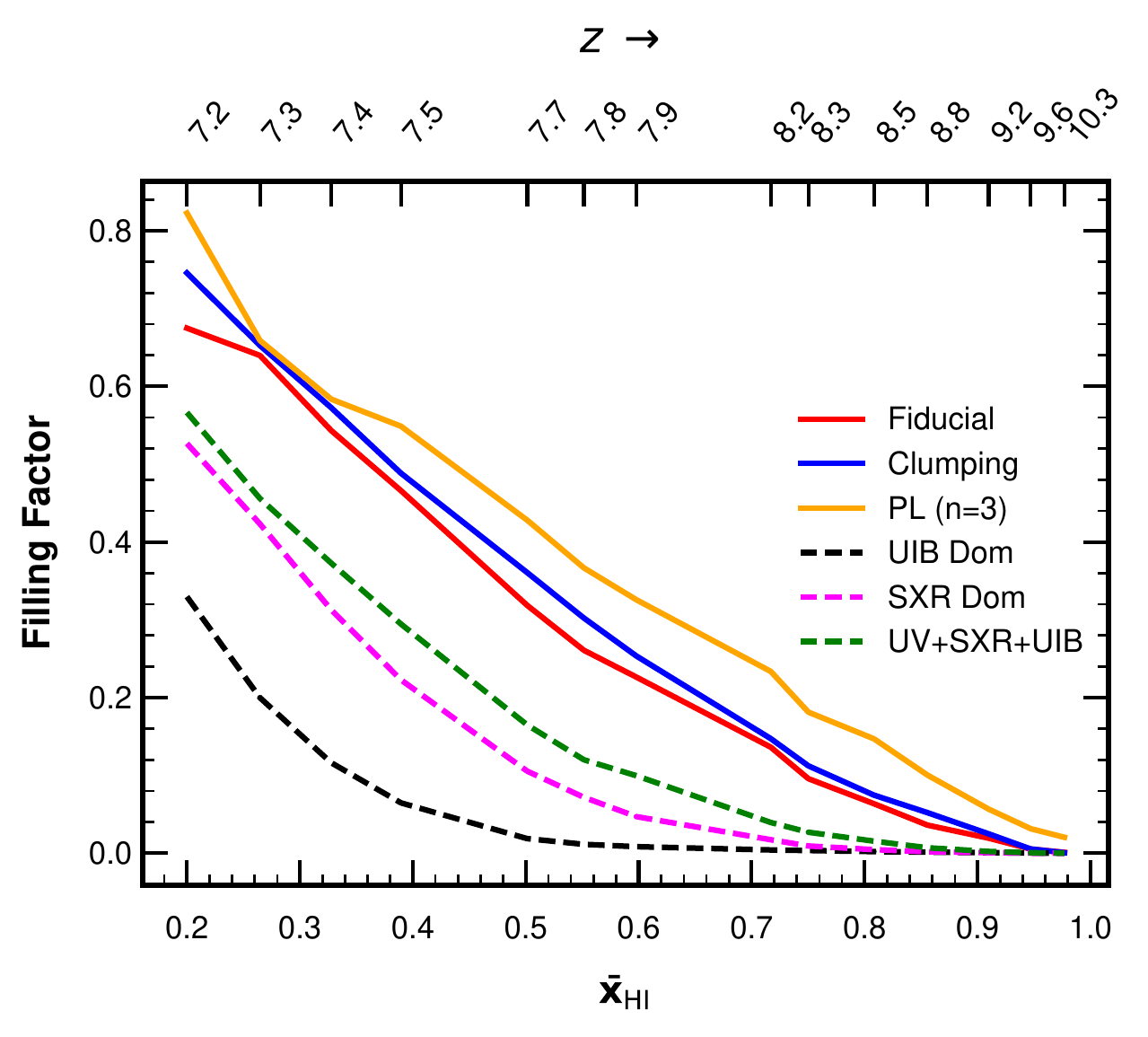}
\caption{{\bf Left panel:} Reionization history. Global neutral fraction as a function of redshift. All six reionization scenarios considered here follow this same reionization history.
{\bf Right panel:} Filling Factor (FF) of ionized hydrogen is plotted against the $\xhi$ for all reionization scenarios. The contribution of FF  is significant in the entire neutral fraction range for scenarios like fiducial, clumping, PL  (n=3) while it become significant at the later stage of reionization in the scenarios like UIB dom. and SXR dom.
}
\label{fig:reion_history}
\end{figure}

In the reionization scenarios considered here, the proportionality constant $N_{{\rm ion}}$ was tuned such that all of them have the same mass averaged neutral fraction ($\xhi(z)$) at a given time/redshift, i.e. they follow the same reionization history. This is illustrated in the left panel of figure \ref{fig:reion_history} which shows the evolution of the neutral fraction with redshift for all of the scenarios. Since different scenarios follow the same $\xhi(z)$, we discuss our results a functions of $\xhi$ instead of redshift. This will help us in understanding how the 21-cm topology/morphology is related to the neutral fraction in different reionization scenarios. 

Note that despite $\xhi$ being same for all scenarios at a given redshift, the total volume of the ionized regions may not be the same in the different scenarios, as demonstrated in the right panel of figure \ref{fig:reion_history}. For a given $\xhi$ (or redshift), UIB dominated scenario results in least FF whereas PL ($n=3$) produces highest FF. Since in the UIB dominated scenario, the hard x-ray photons can escape to longer distances and can effectively produce a uniform ionizing background. This leads to an so called `outside-in' reionization, where the low density regions ionize first and then the high density regions follow the suit. Therefore, in this scenario a large volume will be partially ionized in contrast to PL or Fiducial scenarios. In the Fiducial, Clumping or PL scenarios, the reionization will be `inside-out' in nature, where UV radiations from the collapsed halos escapes to longer distances only after they have substantially ionized their local IGM. The SXR Dom and the UV+SXR+UIB lie somewhere in between the `inside-out' and `outside-in' scenarios.

\section{Methods}
\label{sec:methods}
\subsection{Percolation}
\label{sec:pc}

\begin{figure*}
\hspace*{-2mm}
\subfigure[$\xhi=0.80$, before percolation]{
    \includegraphics[width=0.33\textwidth]{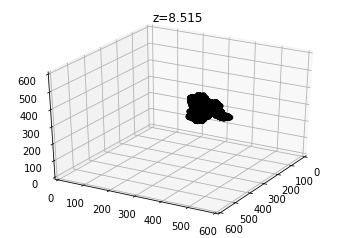}}\hspace*{-2mm}
\subfigure[$\xhi=0.75$, onset of percolation]{
     \includegraphics[width=0.33\textwidth]{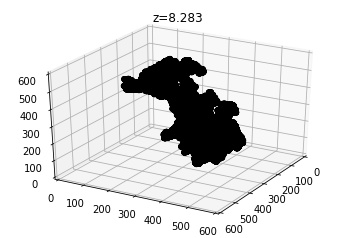}}
     \hspace*{-2mm}
\subfigure[$\xhi=0.72$, just after percolation]{
     \includegraphics[width=0.33\textwidth]{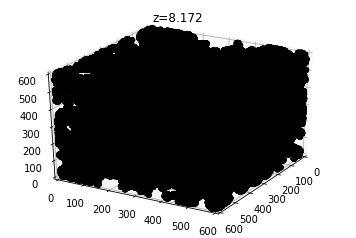}}
\caption{The largest ionized region in the fiducial model has been shown within the simulation volume in the three panels at three neutral fractions, $\bar{x}_{\rm{H}{I}} =0.80, ~0.75,~0.72$ from left to right, characterising well before, just before and just after percolation. At percolation, the largest regions grows abruptly for minuscule changes in the $\xhi$, as seen in the middle to right panel. These plots are shown in grid units of the simulation which has $(600)^3$ grid points and a physical size of $(714.28 {\rm Mpc})^3$.} 
    \label{fig:sim_fidu}
\end{figure*}

When the first luminous objects form at the cosmic dawn (near $z \sim 15$ \cite{choudhury2006}) they start to ionize their surrounding neutral hydrogen field. As reionization progresses, these small ionized hydrogen bubbles grow both in size and number and slowly they start to overlap too. But at some point in time, depending on the reionization scenario, these bubbles suddenly coalesce together to form a large connected single ionized region. This abrupt topological change in the ionization field can be viewed as a `phase' transition and we call it {\em percolation transition} \cite{Klypin1993, Yess1996}. In this work, we track the largest ionized region with redshift for different reionization scenarios. We identify the onset of percolation in a reionization scenario, when the largest ionized region stretches from one face of the simulation box to the opposite face. The largest ionized region is then infinitely extended due to periodic boundary condition of our simulation volume. This is illustrated in figure \ref{fig:sim_fidu} for the fiducial model. The largest ionized region is shown at three stages, well before, just before and just after percolation transition in the panels from left to right respectively. One can see that the largest region grows rapidly from the middle panel to the right one with minute change in the neutral fraction at percolation. Just after percolation, as shown in the right panel, it extends through out the space in all directions.
 
How the percolation transition happens in different reionization scenarios can be studied more comprehensively using the Largest Cluster Statistics (LCS) \cite{Yess1996, Klypin1993, Sahni:1996mb} defined as ,
\begin{equation}
{\rm LCS}=\frac{\rm volume ~of ~the ~largest ~ionized ~region}{\rm total ~volume ~of ~all ~the ~ionized ~regions}\;. \label{eq:lcs}
\end{equation}
Therefore, LCS of the ionized region can be regarded as the fraction ionized volume residing inside the largest ionized region. At percolation transition, we expect that LCS increases sharply with filling factor or $\xhi$. But when and how percolation  takes place in an ionized hydrogen field depend on the reionization process itself. Therefore, we study the evolution of LCS (with FF and $\xhi$) for different reionization models to understand the effects of different ionization processes on the percolation transition and eventually to distinguish the models.

\subsection{Minkowski functionals and Shapefinders}\label{sec:Minkowski}
We complement the percolation analysis with the study of morphology of individual ionized regions using Minkowski functionals (MFs). 
A closed two dimensional  surface has the following four Minkowski functionals \cite{Mecke:1994ax} -- 
\begin{enumerate}
 \item Volume enclosed: $V$,
 
 \item Surface area: $S$,
 
 \item Integrated mean curvature (IMC):
 \begin{equation}
  C=\frac{1}{2} \oint \left(\frac{1}{R_1}+\frac{1}{R_2} \right) dS \;,
 \end{equation}
 
  \item Euler characteristic (or Gaussian mean curvature): 
  \begin{equation}
  \chi=\frac{1}{2\pi} \oint \frac{1}{R_1 R_2} dS\;.
  \end{equation} 
\end{enumerate}
Here $R_1$ and $R_2$ are the two principal radii of curvature at a point on the surface. Euler characteristic (the fourth Minkowski functional) measures the topology of a surface and it can be further expressed in terms of the {\em genus} (G) 
of the surface as follows,
\begin{equation}
 G=1-\chi/2 \equiv {\rm (no.~ of~ tunnels)}-{\rm (no.~ of~ isolated~ surfaces)}+1\;.
\end{equation}
For example, an isolated closed surface with $N_c$ cavities and $N_t$ tunnels passing through it would have genus$\,=N_t-N_c$.

The ratios of these Minkowski functionals are introduced as `Shapefinders' in \cite{Sahni:1998cr} to assess the shape of an object (such as a cluster or a void). In three dimensions, we have three Shapefinders, namely,
\begin{align}
 & {\rm Thickness:}~T =3V/S\;, \label{eq:T} \\
 & {\rm Breadth:}~B =S/C\;, \label{eq:B} \\
 & {\rm Length: }~L =C/(4\pi)\;. \label{eq:L}
\end{align}
All of the Shapefinders -- $T, B, L$ -- have dimension of length, and they are good measures of the extensions of an object in 3-dimensions, hence the name -- thickness, breadth and length\footnote{One finds $T\leqslant B \leqslant L$ in general. However, if the natural order is violated for a region, we choose the largest Shapefinder as $L$ and the smallest one as $T$ to restore the order. In the some rare cases a region may have $C < 0$. We redefine $C \to |C|$ to ensure that all the Shapefinders are positive in those cases.\label{foot:order_SF}}.
The Shapefinders are defined to be spherically normalized, i.e. $V=(4\pi/3) TBL $. 
 
Using the Shapefinders one can further determine the morphology of an object (such as ionized regions), by means of the following dimensionless quantities
\begin{align}
&{\rm Planarity:}~ P =\frac{B-T}{B+T}\;,\label{eq:P} ~~\\ 
&{\rm Filamentarity:}~ F =\frac{L-B}{L+B}\;. \label{eq:F}
\end{align}
As the names suggest, $P$ and $F$ characterize the `planarity' and `filamentarity' of an object and $ 0 \leq P,F \leq 1$ by construction. 
A sphere will have $P \simeq F \simeq 0$, while $P \sim F \sim 1$ for a ribbon. A planar object (such as a sheet) $P \gg F$ whereas the reverse is true for a filament which has $F \gg P$.
In the context of this work, studying Shapefinders of ionized regions at different stages of reionization, together with percolation analysis, would shed light on the evolution of geometry, morphology and topology of the ionization field. 

\subsection{The SURFGEN2 code}
To assess the shape of the ionized regions in terms of Shapefinders, we employed SURFGEN2, which is a more advanced version of SURFGEN algorithm, originally developed by \cite{Sheth:2002rf,Sheth:2005ys, Sheth:2006qz} for studying the largest scale structure of the universe. The improvements in SURFGEN2 have been explained in \cite{Bag18,bag19} in more details. 

\begin{itemize}
\item The code first identifies all the isolated ionized regions, following the definition $\rho_{\rm HI}({\bf x})=0$, in the simulated $\rhi$ field using the `Friends-of-Friends' (FoF) algorithm. Note that, SURFGEN2 finds the regions consistent with periodic boundary conditions.
 
 \item Next, the code models the surface of each ionized region using the advanced `{\em Marching Cube 33}' triangulation scheme \citep{mar33}, which circumvents the issues associated with the original Marching Cube algorithm \citep{marcube}. 
 
 \item Finally, SURFGEN2 determines the Minkowski functionals and Shapefinders of each 
ionized region separately from the triangle vertices that we have from the previous step.

\item We follow the above steps separately at every stage of all the reionization scenarios considered here to determine the shape statistics of ionized regions in those different reionization models.

\item Note that, by tracking the largest ionized region, we can compute the LCS using \eqref{eq:lcs} and study its evolution in different reionization scenarios.
\end{itemize}

\section{Results}
\label{sec:results}

\begin{figure*}
    \includegraphics[width=\linewidth]{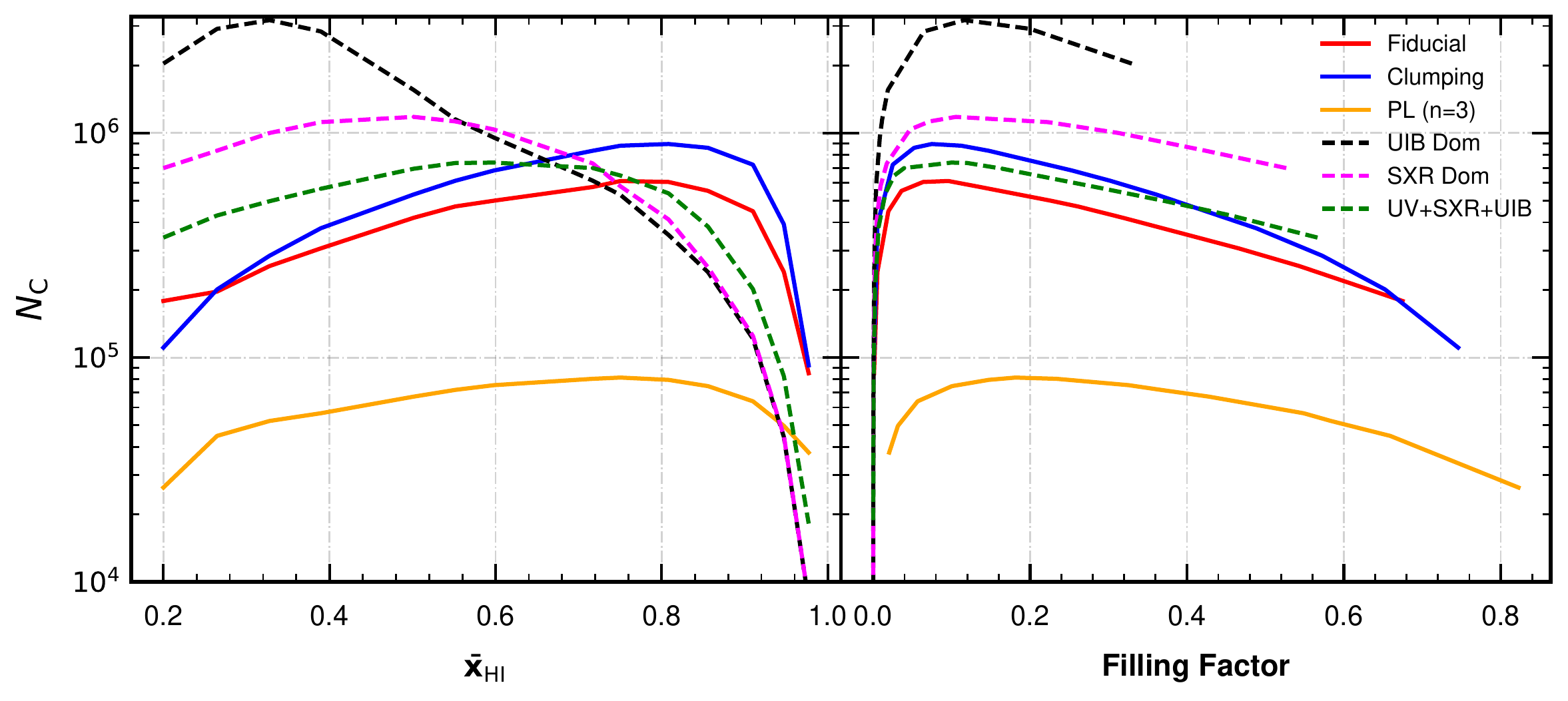}\par 
    
    \caption{The number of ionized regions (\nc) for different reionization models has been shown against the neutral fraction ($\xhi$) and the ionized filling factor (FF) in the left and right panel respectively. Note in the right panel that different models go up to different filling factor by the end of the redshift range we consider in this work.}
    \label{fig:NC_FF}
\end{figure*}

\subsection{Distinguishing different reionization scenarios using cluster statistics and percolation}
\label{sec:cluster_stat}
The {\em cluster statistics} can be very useful in order to distinguish different reionization models, particularly the two fundamental scenarios -- inside-out and outside-in reionization. 
First we focus on the number of ionized regions (\nc) which has been plotted against the neutral fraction ($\xhi$) and the filling factor (FF) respectively in the left and right panels of figure\,\ref{fig:NC_FF} for different reionization scenarios. As reionization progresses, $\xhi$ decreases from unity in the left panel whereas the ionized filling factor increases in the right panel. At the beginning of reionization, \nc\ increases with time as new ionized pockets start to appear as evident from both panels. On the other hand, at the advanced stages of reionization, ionized regions overlap extensively and the number of ionized regions decrease. Therefore, somewhere in between, \nc\ exhibits a maximum in all the reionization models as illustrated in both the panels. However, the way \nc\ reaches to its maximum is different for different reionization scenarios (the percolation transition takes place near the maxima of the respective models, as we would find out below). For example, UIB dominated model (shown in black colour) has the highest number of ionized regions whereas \nc\ in the PL ($n=3$) model (shown in orange colour) is the smallest throughout. Interestingly, as demonstrated in the left panel, the maxima in \nc\ occur at different $\xhi$ values for different models e.g. at early phases for PL $(n=3)$, Fiducial, Clumping models while at late phases for the UIB dominated model (others in between). In contrast, the maxima in \nc\ for different models are  located at similar values of ionized filling factor, $FF \sim 10\%$, as shown in the right panel. Note that different models have different filling factor by the end of the redshift range we consider in this work.

\begin{figure*}
    \includegraphics[width=\linewidth]{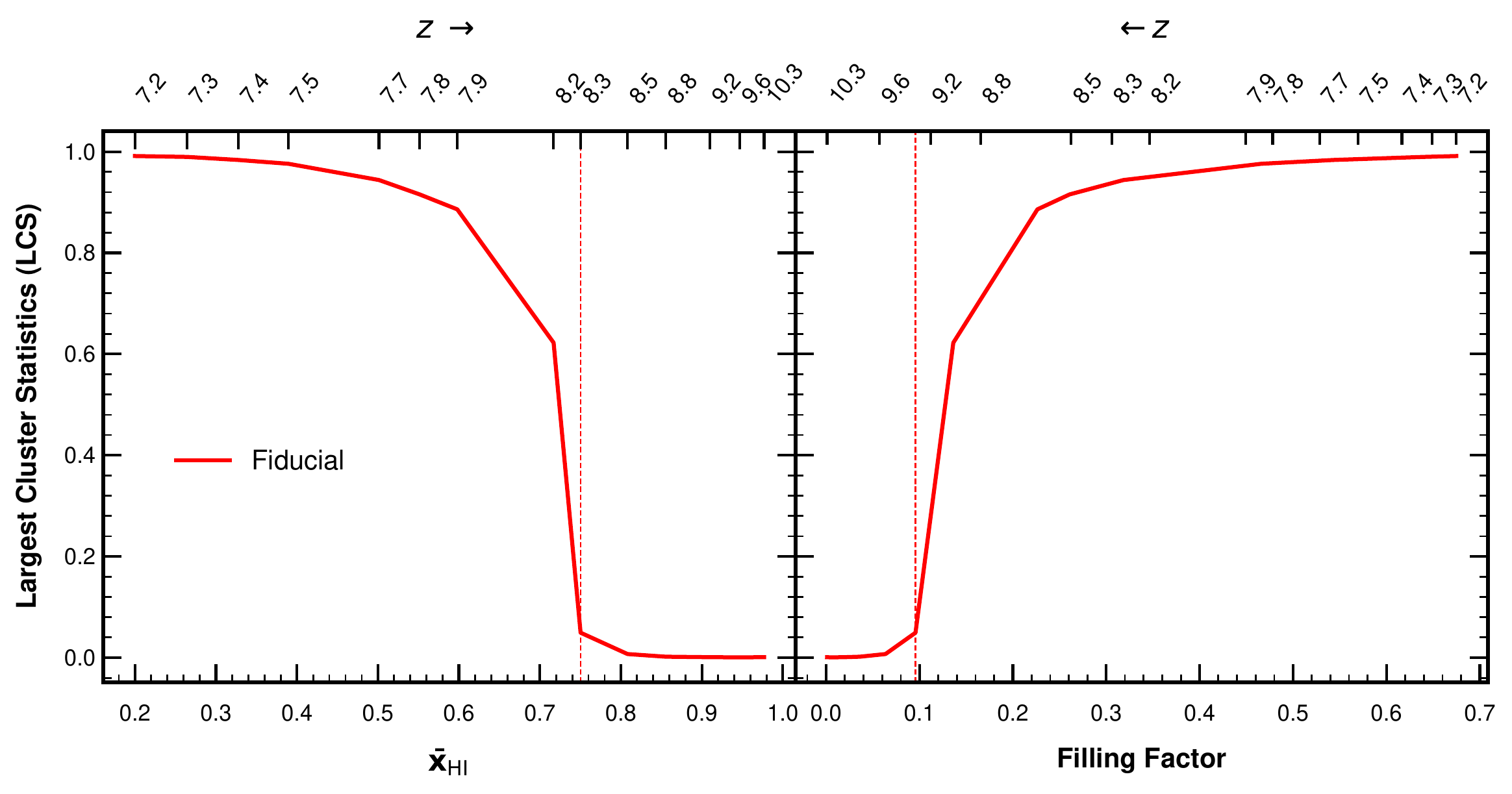}\par 
    
    \caption{The left and right panels show LCS of the ionized hydrogen in the fiducial model as a function of the neutral fraction ($\xhi$) and the filling factor (FF) respectively. In both panels, the vertical dashed line represents the onset of percolation where LCS rises abruptly in this reionization scenario.}
    \label{fig:fidu_LCS_FF}
\end{figure*}

Next, we analyse the percolation (explained in section \ref{sec:pc}) by following the Largest Cluster Statistics (LCS) at multiple $\xhi$. Let us first focus on the results for the fiducial reionization scenario. Figure \ref{fig:fidu_LCS_FF} shows the evolution of LCS with $\xhi$ and filling factor (FF) in the left and right panel respectively. The vertical dashed lines in both panels show the onset of percolation when the largest ionized region abruptly grows in all directions and extend throughout the simulation volume. During the percolation transition, a very little change in $\xhi$ or FF results in a sharp rise in LCS as illustrated in the respective panels. Note that percolation transition itself can be defined through this formally discontinuous growth in LCS \citep{Klypin1993}. We find that percolation transition takes place for the fiducial model at $\xhi \approx 0.75$ and $\ff \approx 9.6 \%$ (corresponds to $z \approx 8.1$) which are consistent with the earlier findings in the literature \citep{Iliev2006, Chardin2012, Furlanetto16, Bag18} where slightly different models were considered. Post percolation, most of the individual ionized regions are assimilated into one large ionized region and the LCS quickly saturates near unity. 

\begin{figure*}
    \includegraphics[width=\linewidth]{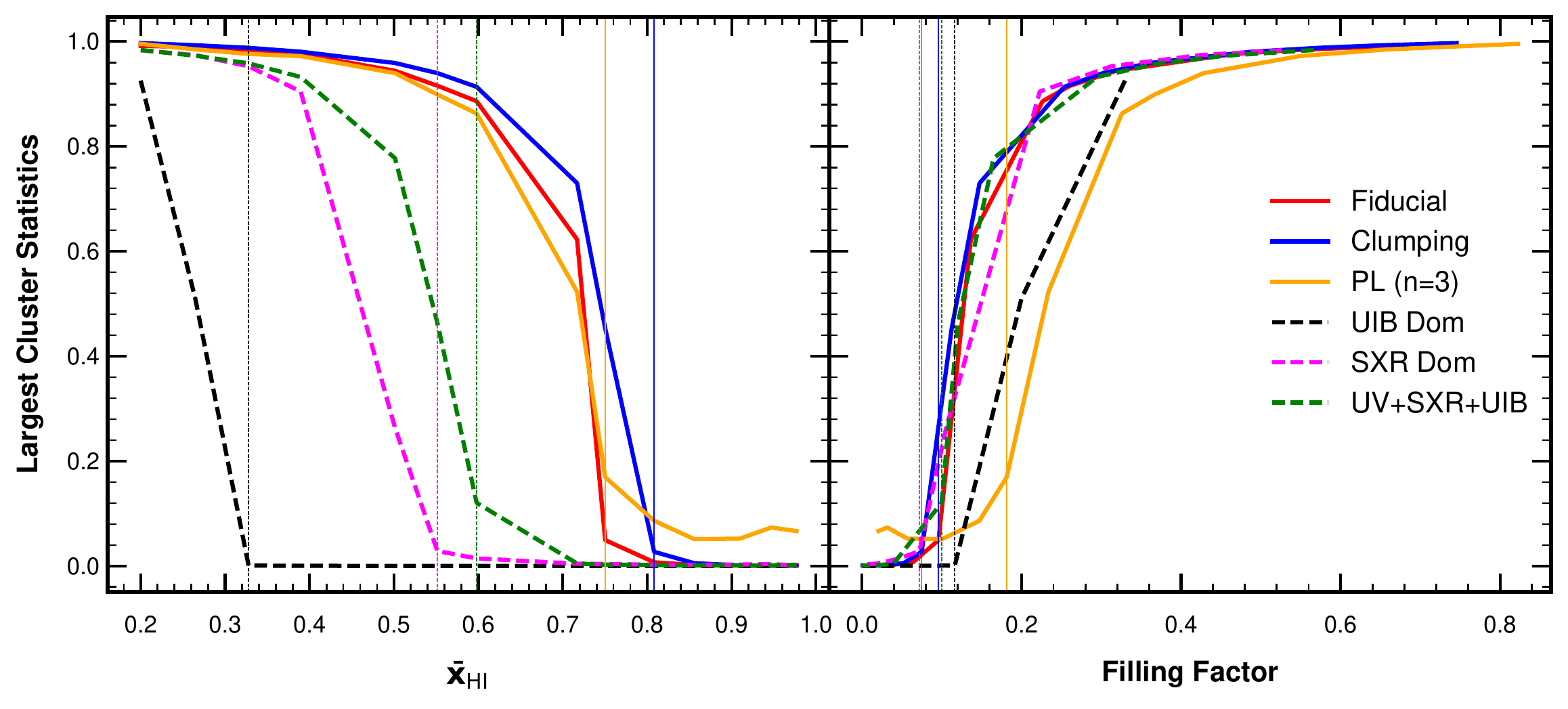}\par 

    \caption{ 
    The percolation transition has been compared for all six reionization models in terms of LCS. The left panel shows the LCS as a function of $\xhi$  whereas the right panel illustrates how LCS evolves with the filling factor (FF) for the different reionization scenarios. In both panels, the vertical lines represent the percolation transitions in the reionization models with corresponding colours. }
    \label{fig:comb_LCS}
\end{figure*}

Next, we compare the percolation transitions in different reionization scenarios in Figure \ref{fig:comb_LCS}. The left and right panels show the evolution of LCS with $\xhi$ and ionized filling factor (FF) respectively for different source models. The onset of percolation in these models is marked by the vertical lines with respective colours in both panels. We find that the LCS exhibits the sharp increase at the percolation transition for all the models, similar to what we observe for fiducial model above. However, the ionized hydrogen in different scenarios percolates at different values of $\xhi$ but at only slightly different FF values. The first two columns of the table \ref{tab:table_all} present the critical neutral fractions and the ionized FF where percolation take place for each reionization source model.

Focusing on the LCS vs $\xhi$ curves in the left panel of Figure \ref{fig:comb_LCS}, one can get a wealth of information regarding the nature and properties of the source models or reionization scenarios. 
\textit{(1)} Firstly, in all reionization scenarios the ionized hydrogen percolates at very different global neutral fraction values. This clearly demonstrates that percolation is strongly connected to the 21-cm topology. \textit{(2)} Although we can see a sharp ascent in the LCS for all reionization scenarios at the percolation transition but the shape of the LCS curves is different for different scenarios. 
For example, in PL $(n=3)$ model, the large ionized regions could easily connect and eventually percolate at a relatively earlier stage of reionization. In contrast, the percolation transition in UIB dominated model takes place at an advance stage of reionization because of the ionized regions being relatively  finer and the filling factor being low.
In summary, ionized hydrogen in the scenarios like fiducial, clumping, PL (n=3),  percolates at higher neutral fractions because of the inside-out reionization in these cases. 
On the other hand, in the outside-in reionization scenarios, like UIB dominated, SXR dominated etc., percolation takes place at much lower neutral fractions once the filling factor becomes significant. In view of the right panel of figure \ref{fig:reion_history}, one finds that the filling factor in these scenarios could grow substantially only after the respective percolation transitions. \textit{(3)} For PL $(n=3)$ (shown by the orange curve), LCS starts with a larger value at the beginning of the EoR (i.e. before percolation) compared to the other models. This is the manifestation of the fact that this reionization scenario creates larger (in size) but fewer ionized regions as compare to other scenarios, as shown in figure \ref{fig:NC_FF}. \textit{(4)} The clumping model shows percolation at very early stage of the EoR (nearly $z \approx 8.515$) since this is the only scenario where non-uniform recombination has been taken into the account. Although this {\em nonphysical} scenario could not be distinguished from the others using the percolation analysis, the shape diagnostic tools -- Shapefinders -- can isolate it from the rests, as we demonstrate below.

From the LCS vs FF curves, presented in the right panel of figure \ref{fig:comb_LCS}, we observe similar characteristics. However, percolation transitions in all the models take place in a narrow range of ionized filling factor, $\ff \in (10-20 \%)$. Therefore, we find that LCS vs FF curves are not as sensitive to the reionization scenarios as the LCS vs $\xhi$ curves, making the former curves not quite suitable for distinguishing reionization models.

We note that the cluster statistics ideally should be accompanied by the uncertainties in order to distinguish various source models. A proper error estimation requires analysing a large number of realizations of the simulations for each scenario at each redshift. Simulating many high-resolution realizations for every case is beyond the scope of the present paper. Nevertheless, we crudely estimate the errors in appendix \ref{app:uncertainties} on filling factor, number of ionized regions per unit volume and LCS  by dividing the large primary simulation volume in 8 equal smaller volumes those are themselves large enough to obey the cosmological principles. 
The uncertainties we find in LCS are reasonably tight, larger for inside-out models as compared to the outside-in models. Remarkably, we find that LCS with the tight errorbars is very much suitable for distinguishing the source models, especially between the inside-out and outside-in scenarios with high certainty.   

We have further checked the consistency of our findings from the LCS analysis by comparing the results with the Bubble Size Distribution (BSD) in different scenarios. The BSD shows a bimodal distribution where the largest ionized bubble is largely separated from the bulk after percolation for all reionization scenarios which is consistent with the LCS evolution since friends-of-friends is used to find the ionized regions in both approaches.

\begin{table}
    \centering
    \begin{tabular}{ |P{3.0cm}||P{1.4cm}|P{1.2cm}|P{1.6cm}|P{2.35cm}| }
 \hline

     \textbf{Reionzation scenarios} & \textbf{Critical $\bar{x}_{\rm{H}{I}}$} &
     \textbf{$FF_C$} &
     \textbf{Planarity} & \textbf{Filamentarity}\\
 \hline \hline
 Fiducial & 0.75 & 0.096 & 0.162 & 0.998\\
 \hline
 Clumping & 0.81 & 0.075 & 0.858 & 0.914\\
 \hline
  PL  (n=3) & 0.75 & 0.181 & 0.041 & 0.988\\
 \hline
 UIB Dominated & 0.33 & 0.116 & 0.269 & 0.979\\
 \hline
 SXR Dominated & 0.55 & 0.072 & 0.177 & 0.993\\
 \hline
 UV + SXR + UIB & 0.59 & 0.099 & 0.104 & 0.999\\
 \hline
\end{tabular}\\

\caption{All reionization scenarios are compared at percolation in terms of the critical neutral fraction, planarity and filamentarity of the largest ionized region. \textit{$FF_C$} corresponds to the Filling Factor at which percolation transition takes place.}
    \label{tab:table_all}
\end{table}

\subsection{Shape of the largest ionized region}
\label{sec:SF_lir}
\begin{figure*}
    \centering
    \includegraphics[scale=0.5]{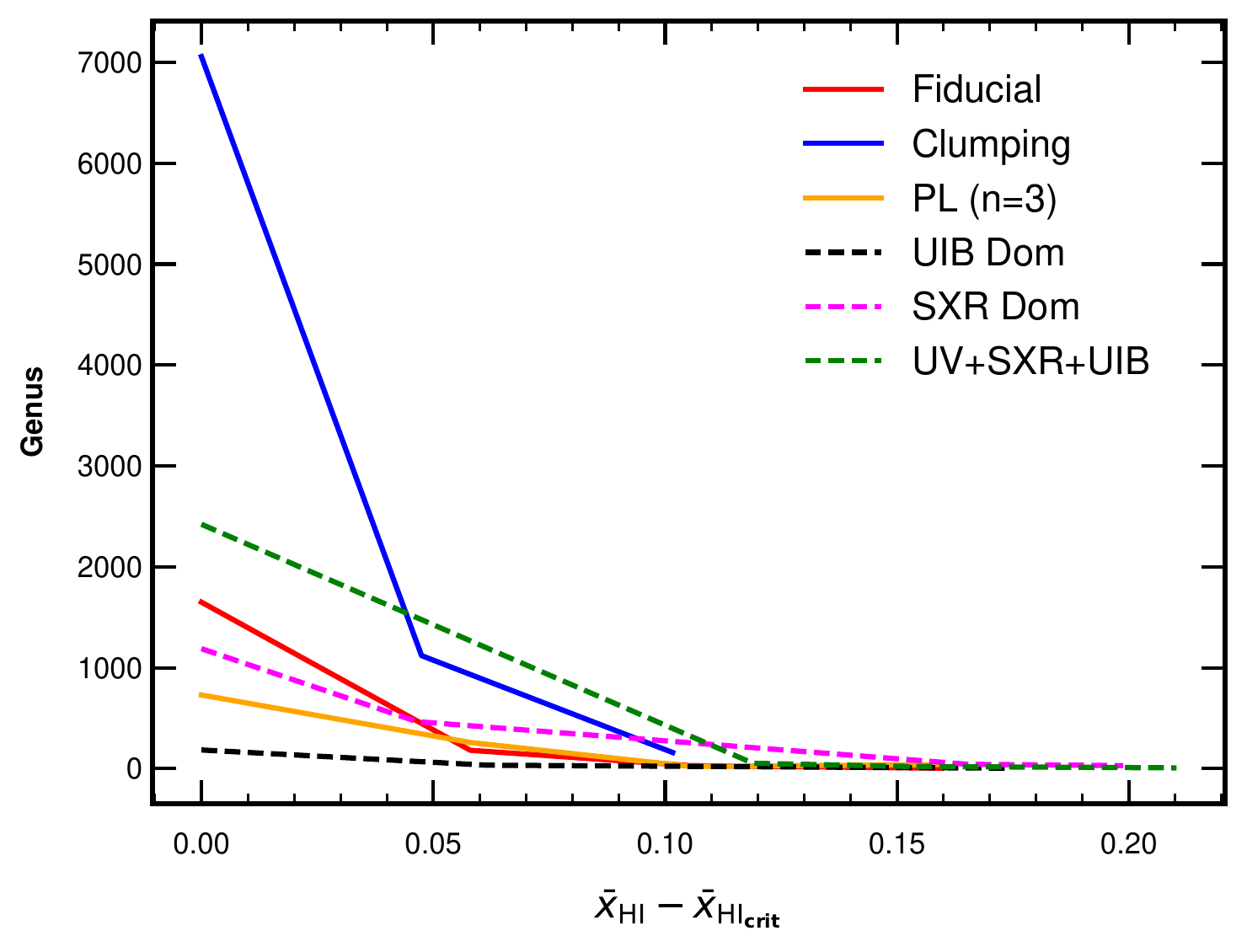}
    \includegraphics[scale=0.5]{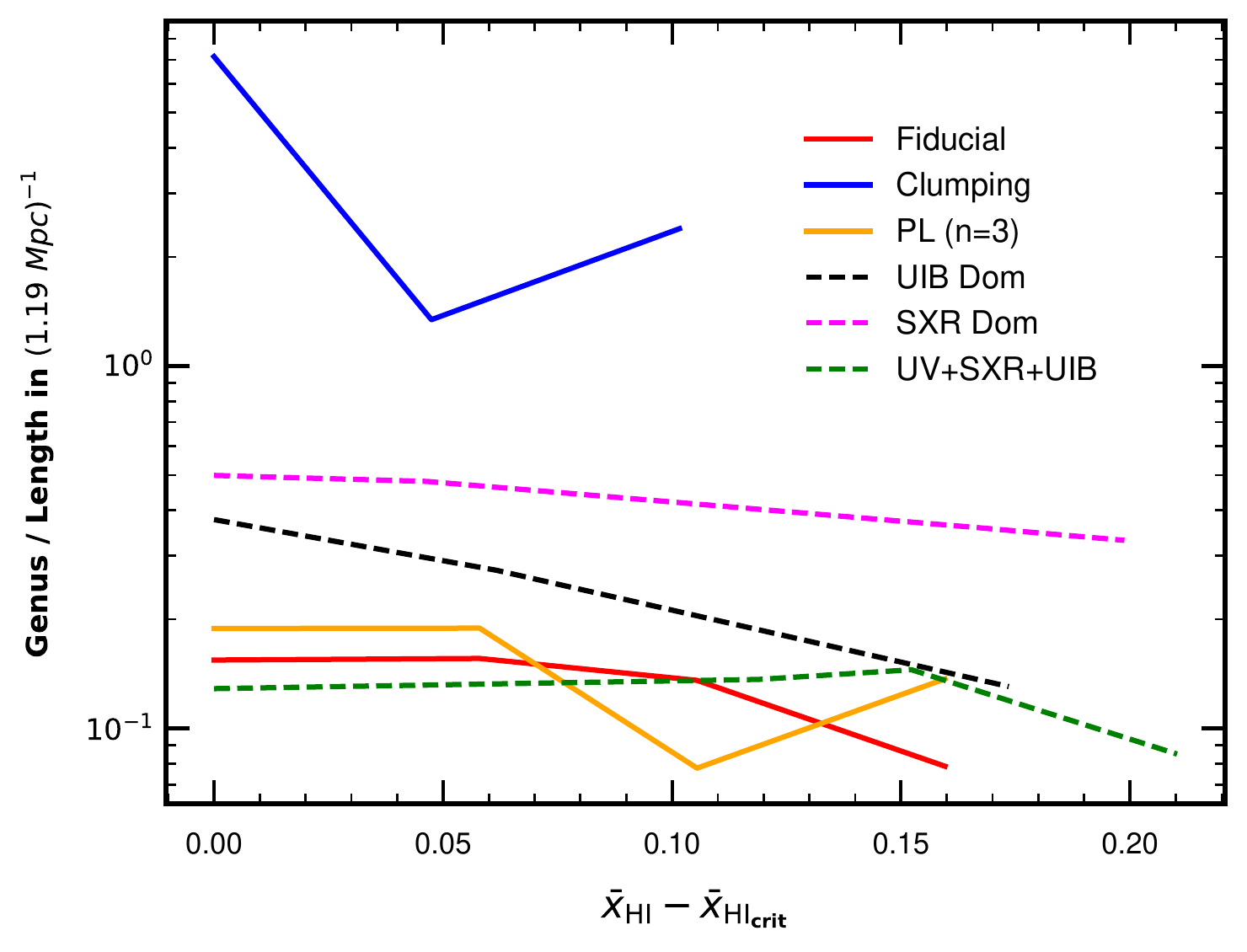}
    \caption{Genus of largest ionized region is compared in the left panel for all reionization scenarios up to the percolation transition in linear scale ($\xhi=\xhic$). The right panel which shows the genus per unit `length' in log-scale for the largest regions illustrates that for all scenarios the genus is almost proportional to the length except for the clumping model. }
    \label{fig:lir_genus}
\end{figure*}

\begin{figure}
    \centering
    \includegraphics[scale=0.58]{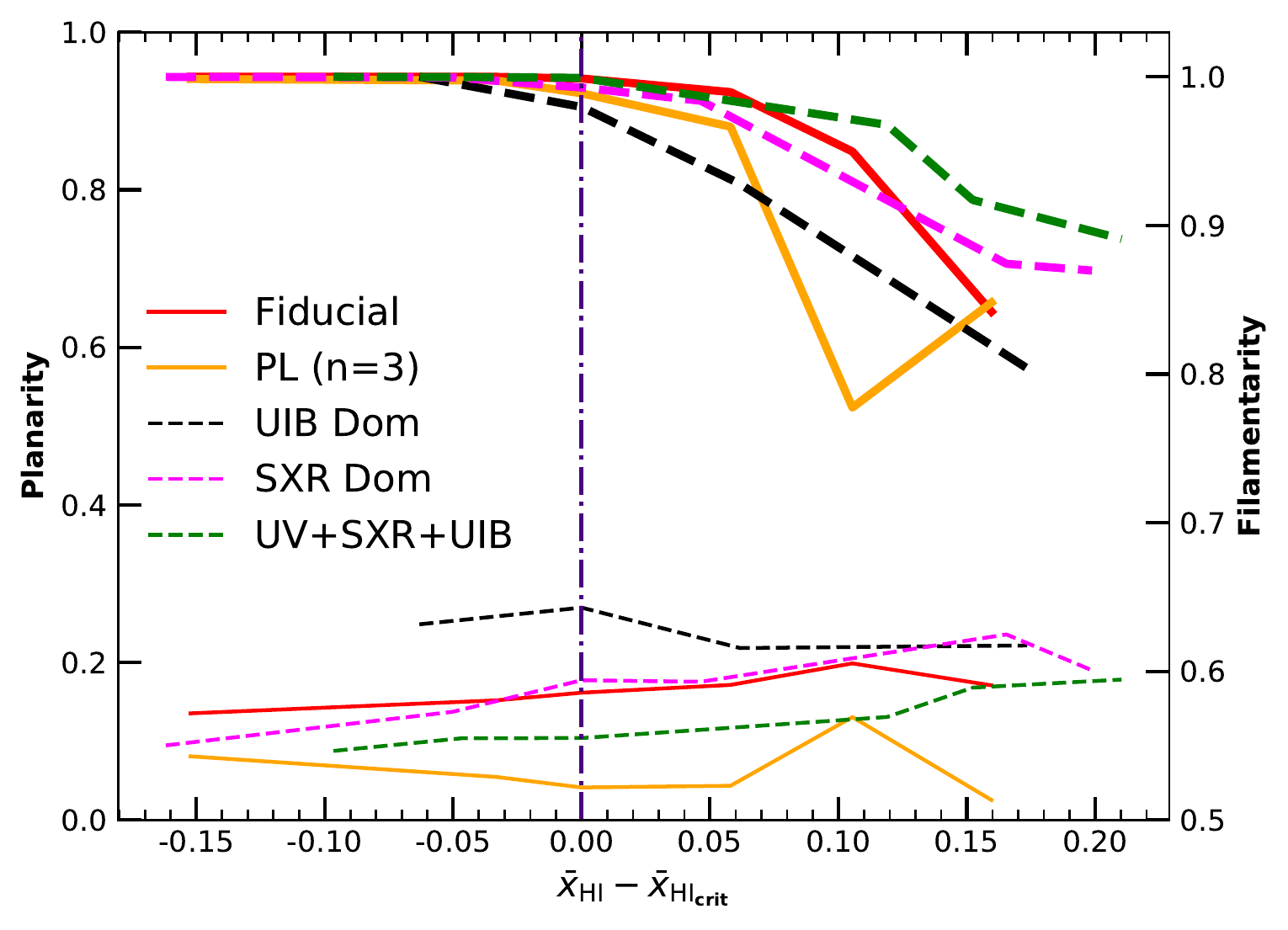}
    \caption{Planarity and filamentarity of the largest ionized region in different models have been compared near the respective percolation transitions. The solid and dashed curves represent the reionization scenarios following inside-out and outside-in reionization respectively. The curves shown with thick lines (solid and dashed) represent the filamentarity and the curves with thin lines represent the planarity of the largest ionized region in different models. Note that percolation transition takes place at different $\xhic$ for different models, thus at $\xhi-\xhic=0$ (shown by the dashed vertical line) for all the models. }
    \label{fig:PF_LC}
\end{figure}

In this section, we discuss the evolution of the topology, morphology and shape of the largest ionized region based on the Minkowski functionals and the Shapefinders (see the section \ref{sec:Minkowski}). The left panel of figure \ref{fig:lir_genus} illustrates how the topology of the largest ionized region in different reionization scenarios evolves with time. The genus value of the largest ionized region has been plotted against the neutral fraction in different scenarios till the percolation transition\footnote{Because of the periodic boundary condition, the physical shape of a percolating region cannot be defined. Therefore, the genus value of a region just after percolation is also not well defined. However, well beyond percolation when the ionized region covers most of the simulation volume, one can reliably estimate the genus value in terms of per unit volume.} in the respective scenarios. Note that we shift the x-axis by the critical neutral fraction (at the onset of percolation) for each model. We observe that as the reionization progresses the largest ionized regions in all the scenarios become more multiply connected and their genus increases. As we find out later, the `length' (the 3rd Shapefinder) of the largest region also increases with reionization. The right panel of figure \ref{fig:lir_genus} shows the evolution of the genus per unit `length' for the largest regions in these scenarios. In all the scenarios, except the clumping model, genus per unit `length' remains stable. 

It is evident from the figure \ref{fig:lir_genus} that the genus of the largest region in the clumping model is much higher (by several orders) than that in the other scenarios despite the fact that ionized hydrogen percolates earlier in time in the clumping model. Because of the aggressive non-uniform recombination in the clumping scenario, there remain many pockets of neutral hydrogen which tunnels through the ionized regions giving rise to the high genus values. Most of these tunnels are found to be negatively curved and hence they lead to decrease in the overall integrated mean curvature (the 3rd MF). This in turn breaks the natural order of the Shapefinders, i.e. $T<B<L$ and results in higher planarity value that, one can argue, is not physical. Therefore, we do not include the Shapefinders results for the clumping model in the rest of the main paper, if not mentioned otherwise. We analyse the results for the clumping model separately in the appendix \ref{app:clumping} where we show how this `non-physical' high planarity can distinguish the clumping model from the rests.

The planarity ($P$) and filamentarity ($F$) of the largest ionized regions in different reionization scenarios (except the clumping model) have been shown in figure \ref{fig:PF_LC} near the percolation transitions in the ionized segment for the respective models. The filamentarity of the largest ionized regions in all the models increases as reionization progresses and it reaches to almost unity on approaching the percolation transition. In contrast, the planarity of the largest regions always remains low, $P \lesssim 0.2$ considering all the models. Thus one can conclude that the largest ionized regions in all the reionization scenarios become highly filamentary near the percolation transition, consistent with what \cite{Bag18} found out for a single reionization scenario. The  planarity and filamentarity of the largest ionized region at the onset of percolation have been given in the third and fourth columns of the table \ref{tab:table_all} respectively for all the ionization source models.

\begin{figure*}
    \includegraphics[width=\linewidth]{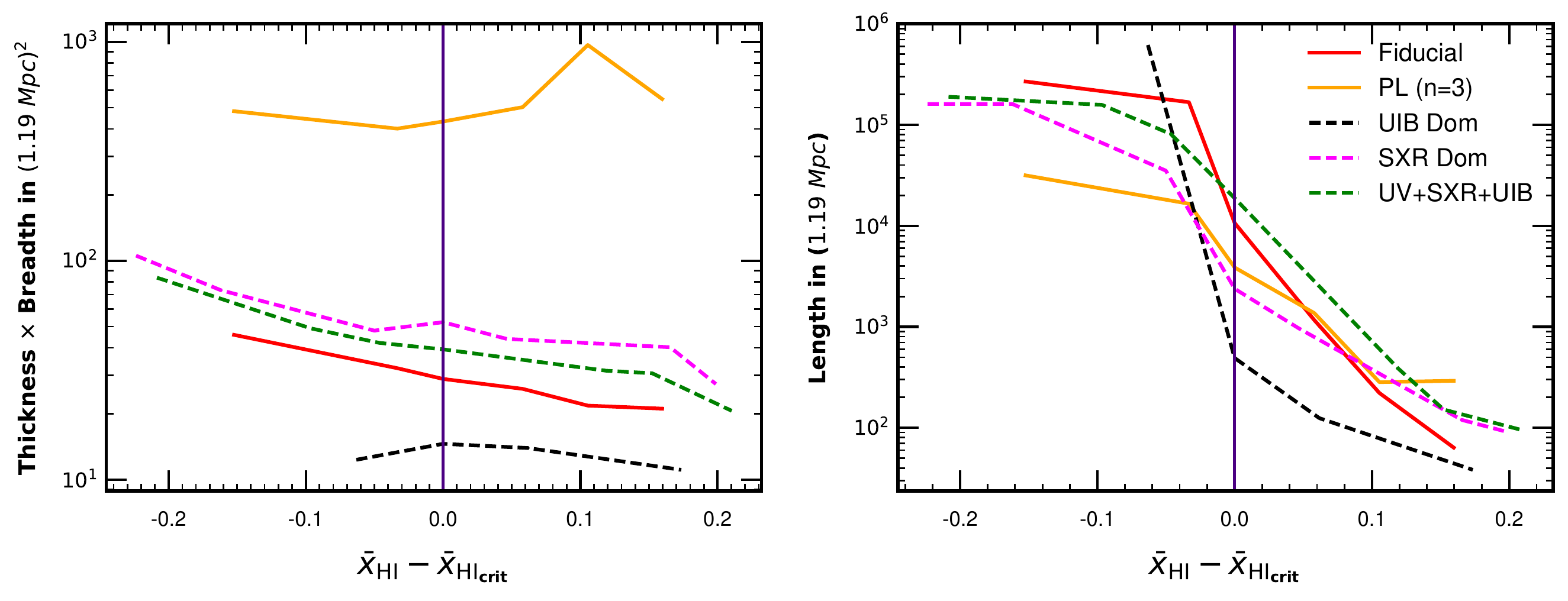}\par 
    
    \caption{This figure illustrates the evolution of the shape of the largest ionized region (LIR) for the five reionization scenarios (excluding the clumping model) near the respective percolation transitions. The left and right panels show the cross-section (estimated by $T \times B$) and the length ($L$) of the LIR against the neutral fraction ($\bar{x}_{\rm{H}{I}} - \bar{x}_{\rm{H}{I}_{crit}}$). Note that $\bar{x}_{\rm{H}{I}_{crit}}$ represents the neutral fraction at the onset of the percolation transition in the ionized hydrogen, it is different for different models as given in table \ref{tab:table_all}.
    }
    \label{fig:TXB_all}
\end{figure*}

For a filamentary object, $T \times B$ can be regarded as the effective `cross-section' (of the filament). Bag et al 2019 \cite{Bag18} pointed out that the  largest ionized region (in their simulation) abruptly grows only in terms of its third Shapefinder , `length ($L$)', during the percolation transition while its `cross-section' remains stable. Here we put this claim to test for different scenarios. The left panel of figure \ref{fig:TXB_all} shows the cross-section  ($T \times B$) of the largest ionized regions in different scenarios against the neutral fraction near the percolation in the respective models. Their lengths have been plotted in log-scale in the right panel. The vertical dashed lines in both panels correspond to the onset of percolation in each model. 

Comparing both panels one can find out that the lengths are in general several orders of magnitude higher than the cross sections. Indeed, we find that the cross-section of largest ionized regions in different models does not increase much during percolation  whereas their length rises sharply. Therefore, the claim of \cite{Bag18} that the largest region grows in-terms of its length only during percolation has been somewhat followed in all the models. One can further observe in the left panel of the figure \ref{fig:TXB_all} that this stable `characteristic' cross-section has the highest value for the PL $(n=3)$ model while the lowest value for the UIB dominated model among the ones we consider in this work. For UIB dominated reionization source model, the ionized bubbles are much smaller in size (and more uniformly distributed in the space) as compared to, say, the fiducial or the PL $(n=3)$ model. Therefore, the inter-connected filamentary largest region arising because of merger of these tiny bubbles in the UIB dominated model also has smaller cross-section as compared to other models. On the other hand, since PL $(n=3)$ produces large ionized bubbles to start with, the largest region also possesses wider effective cross-section.

\subsection{Shape distribution of ionized regions in different reionization scenarios}
In this section we study the shape (together with topology and morphology)  distribution of the individual ionized regions  in the different source models at the onset of the percolation transitions. We divide the regions into 8 volume bins. The errorbars represent the standard deviation which illustrate the scatter of respective quantities in each bin.

\begin{figure*}

    \includegraphics[width=\linewidth]{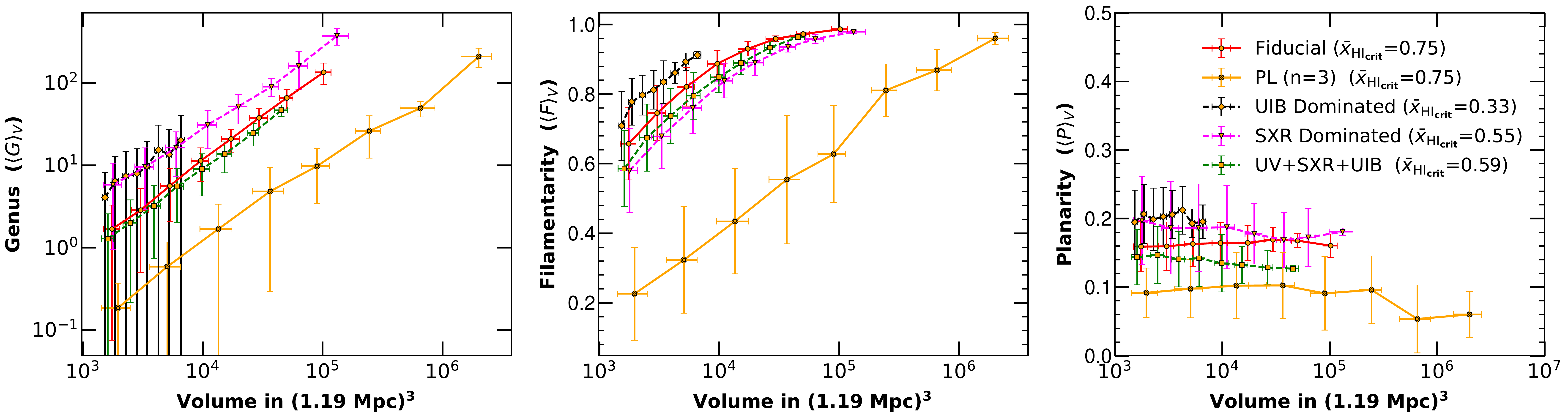}\par 
   
\caption{The figure shows how the shape of the ionized regions at the onset of percolation transition depends on the volume in different reionization scenarios.
The ionized regions at $\xhi=\xhic$ are divided into 8 volume bins (equispaced in log-scale), separately for different reionization scenarios. (Here we ignore the smaller ones with volume below $\sim 100 ~\mpc^3$ as they are too many in number and mostly spherical in shape with trivial topology.)
The three panels, from left to right, show the volume averaged genus ($\langle G \rangle_V$), filamentarity ($\langle F \rangle_V$) and planarity ($\langle P \rangle_V$) values of the ionized regions in different volume bins. Each plot also shows the variation of above mentioned quantities for all reionization scenarios as the error bars depict the standard deviations of these quantities in each bin.}
    \label{fig:PFG}
\end{figure*}

The left most panel of figure \ref{fig:PFG} shows the volume averaged genus values of ionized regions falling in different volume bins for all the source models (excluding the clumping model). For all the models, larger regions are more multiply connected with higher average genus values. The PL (n=3) model produces large ionized regions and the growth of genus with volume is shallower than that in the other models. On the other hand, for the SXR dominated model average genus value increases more rapidly for the higher volume bins. In general we observe the trend that for outside-in models the larger ionized regions tend to be more multiply connected. Note that the UIB dominated model produces small ionized regions, hence the curve is truncated at much lower volume. 

The rest two panels show the volume averaged filamentarity and planarity, respectively, of the ionized regions belonging to different volume bins. For all the models, the ionized regions have very low planarity. But the filamentarity tends to increase with the volume. Interestingly, for the PL (n=3) model the growth of filamentarity with volume is the slowest among all the scenarios we consider here.
On the other hand, the ionized regions in the UIB dominated model have overall higher filamentarity due to their narrower cross-sections. However, we conclude that the large ionized regions in all model tend to be filamentary with high filamentarity and low planarity.

\section{Summary and discussion}
\label{sec:conclusion}
We have studied the percolation transition in the ionized hydrogen by tracing the largest cluster statistics (LCS) in the 3D redshifted 21-cm maps of EoR using a suite of semi-numerical simulations. We compared different reionization scenarios in terms of the geometrical properties of the ionized regions and specifically the way they evolve in the vicinity of percolation transition. We used SURFGEN2 algorithm which implements the shape finding techniques like Minkowski functionals and Shapefinders in conjunction with the percolation analysis to explore the morphology of the 21-cm field. The key points of our work are summarized below:

\begin{itemize}

   \item The number of isolated ionized regions formed by the different reionization scenarios essentially informs about the nature of the undergoing reionization process. Scenario like UIB dominated has the largest value of \nc\ which illustrates the fact that the low dense regions (voids in between the filaments) which are mainly larger in numbers are ionized first and hence obeys outside-in reionization. While on the other hand, small value of \nc\ in scenario like PL ($n=3$) indicates that the highly dense regions are being ionized first and hence it will extensively follow inside-out reionization.

    \item The largest ionized region before percolation transition is almost indistinguishable from other ionized regions for all reionization scenarios. LCS starts to grow at the onset of percolation transition in the ionized hydrogen. Since the evolution of LCS is sensitive to the reionization scenarios, it can be used to distinguish different source models. We can further classify the scenarios in two broad categories, inside-out and outside-in reionization. For the inside-out reionization scenarios (that include the fiducial, PL ($n=3$) and clumping models) percolation transition in the ionized hydrogen takes place at an earlier phase (larger $\xhi$) in contrast to the outside-in scenarios (UIB, SXR dominated and UV+SXR+UIB scenarios) where ionized hydrogen percolates only at low $\xhi$ values, i.e. at advanced stages of reionization. These findings based on LCS are found to be consistent with that from the Bubble Size Distributions.

    \item We also find that the cross section of the largest ionized region in all the scenarios (except for the clumping model) remains stable at the percolation transition when the largest ionized region abruptly grows, mostly in terms of its third Shapefinder -- `length'. Therefore, the filamentary largest ionized region at percolation exhibits a characteristic cross-section, its value depends on the reionization source model. The only exception is the clumping model where the extreme inhomogeneity in the IGM results in a boost in the planarity of the ionized regions (including the largest one) that might not be entire physical. However, one can discern such non-uniform recombination models  by assessing the Shapefinders in the ionized hydrogen. 
    
    \item The genus of largest ionized region for all scenarios increases as the reionization progresses which confirms the multi-connectedness we observe at the later stage of reionization. Except of the unusual behaviour for Clumping scenario, rest of the reionization scenarios show a near linear relationship of genus with length. 
    
\end{itemize} 

To estimate the uncertainty in our geometrical analyses, we ideally need to study many realizations of simulation at each redshift (or $\xhi$) separately for each reionization source model. Such a large number of $N$-body simulations with the fine mass resolution is very expensive and beyond the scope of the present paper. Instead, we crudely estimate the uncertainties (shown by error bars in figures \ref{fig:FF_err} -- \ref{fig:LCS_xhi_err}) by dividing each original simulation volume into eight smaller sub-volumes followed by estimating the variance of various quantities from these sub-volumes. We have discussed this in detail in Appendix \ref{app:uncertainties}.

However, the noise in the observation and the primary beam shape can affect our results further. The partially ionized hydrogen in the presence of observation noise in the 21-cm brightness temperature maps poses additional challenges in defining the ionized hydrogen. Exploring how all these affect our analyses is beyond the scope of the present paper. However, in appendix \ref{app:low_res} we investigate the impact of data resolution on our study by down-sampling the maps with two times coarser resolution. We find that our key results (e.g. the evolution of LCS and the shape of the largest ionized region) are mostly unaffected by the change of resolution. In follow-up, we plan to study the impact system noise and residual foregrounds in the 21-cm maps on our inferences drawn from LCS and Shapefinders of ionized regions and we would also like to explore whether these geometrical tools can be used to put constraints on reionization history via the future SKA observations of the EoR.

\label{sec:discuss}

\section{Acknowledgements}
The authors thank the anonymous reviewer for their constructive comments, which have helped in improving the quality of this article. SB thanks Varun Sahni, Santanu Das for their contributions in developing SURFGEN2 in its initial phase. SM acknowledges financial support through the project titled ``Observing the Cosmic Dawn in Multicolour using Next Generation Telescopes'' funded by the Science and Engineering Research Board (SERB), Department of Science and Technology, Government of India through the Core Research Grant No. CRG/2021/004025. RM is supported by the Israel Academy of Sciences and Humanities \& Council for Higher Education Excellence Fellowship Program for International Postdoctoral Researchers. The entire analysis of the simulated 21-cm maps presented here were done using the computing facilities available with the Cosmology with Statistical Inference (CSI) research group at IIT Indore. 



\appendix

\section{Estimating uncertainties in the LCS using sub-volumes of the simulated signal}
\label{app:uncertainties}

In section \ref{sec:cluster_stat} we distinguish the inside-out reionization scenarios from the outside-in models using cluster statistics and by comparing the percolation transitions in the different reionization scenarios. However, one needs to account for the uncertainties in the respective quantities, especially on the largest cluster statistics (LCS) vs $\xhi$ curves. 
A proper error estimation requires analysing many realizations in all the scenarios, however simulating large number of high-resolution realizations would be extremely computationally expensive and beyond the scope of the present paper. In this section, we crudely estimate the errors on the important quantities by breaking our large $(714 ~\mpc)^3$ simulation volume into 8 equal smaller $(357 ~\mpc)^3$ boxes preserving the resolution. Since the dimensions of the smaller boxes are sufficiently large to satisfy the cosmological principles of isotropy and homogeneity, one can effectively consider the smaller volumes as independent realizations. Therefore, all the cosmological phenomena including the percolation process can be compared between the larger and the smaller simulation volumes. 
Strictly speaking, here we do not follow the periodic boundary condition (PBC) properly. However, since the percolation transition happens abruptly in the whole volume, the effect of PBC can be ignored for this exercise.

\begin{figure}
    \centering
    \includegraphics[scale=0.65]{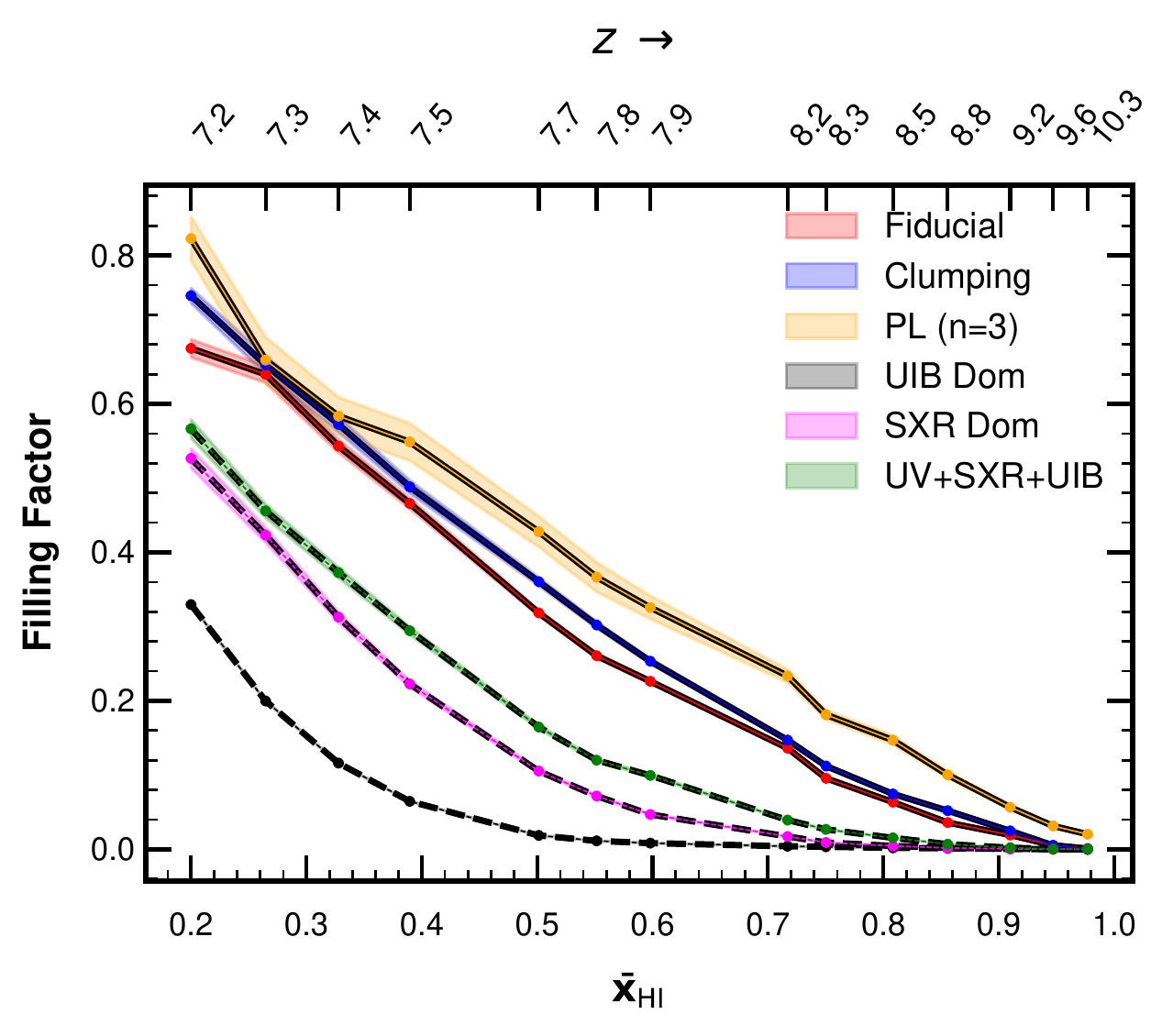}
    \caption{Same as the right panel of figure \protect\ref{fig:reion_history} but with the uncertainties (shown by the shaded regions) estimated using the standard deviation of the filling factor obtained from the smaller boxes analysed separately.}
    \label{fig:FF_err}
\end{figure}

We study the smaller 8 boxes separately and, as expected, we find that the quantities like filling factor, LCS etc are statistically consistent between the larger $(714 ~\mpc)^3$ box and the smaller boxes. This allows us to estimate the uncertainties in various quantities from their standard deviations.
In figure \ref{fig:FF_err}, we show the ionized filling factor against the neutral fraction ($\xhi$) for the different scenarios but this time with the uncertainties calculated using the smaller boxes. The solid and dashed curves represent the results coming from our primary large simulation volume that is in extremely good agreement with the results of the 8 smaller boxes analysed separately. 
The shaded regions with respective colours represent the uncertainty estimated from the standard deviation of the individual results for these smaller boxes. Note that the uncertainty is larger for the inside-out reionization scenarios (maximum for the PL ($n=3$) model) where typically larger ionized bubbles are produced. In contrast, for the outside-in models we get negligible error on the filling factor (minimum for the UIB dominated model) owing to the fact that smaller ionized bubbles formed in these scenarios leading to less variation among different realizations. 

\begin{figure*}
    \includegraphics[width=\linewidth]{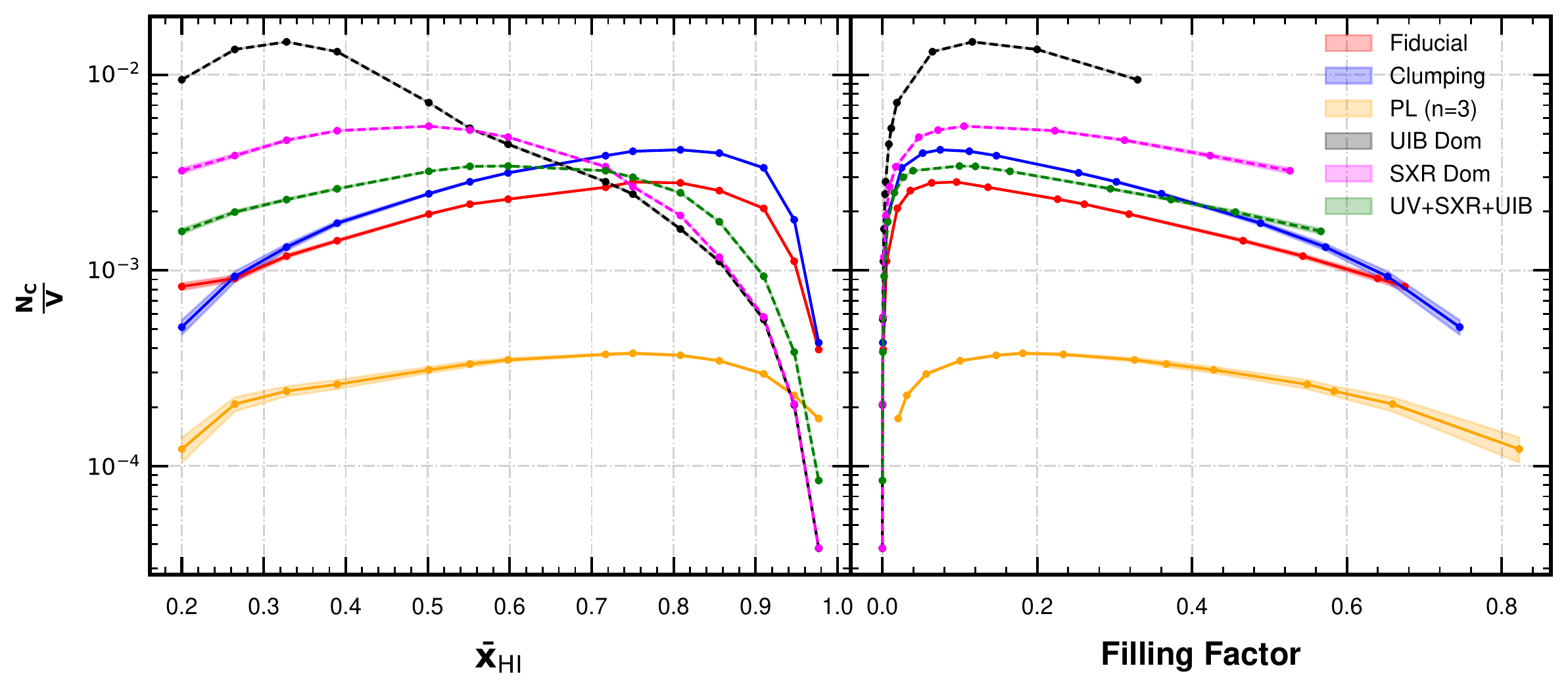}\par 
    \caption{The number of ionized regions per unit volume for different reionization models has been shown against the neutral fraction ($\xhi$) and the ionized filling factor (FF) in the left and right panel respectively. Note in the right panel that different models go up to different filling factor by the end of the redshift range we consider in this work. The shaded regions represent the standard deviations of the results for the smaller boxes studied separately.}
    \label{fig:NcV_err}
\end{figure*}

For a fair comparison, we show the evolution of the number of ionized regions per unit volume ($N_{\rm C}/V$) in figure \ref{fig:NcV_err}; the left and right panels show ($N_{\rm C}/V$) along with the uncertainty as functions of $\xhi$ and filling factor respectively. 
Again we note that the results from our primary large simulation volume is consistent with that from the smaller boxes. The uncertainty is larger for inside-out models.

\begin{figure*}
    \includegraphics[width=\textwidth]{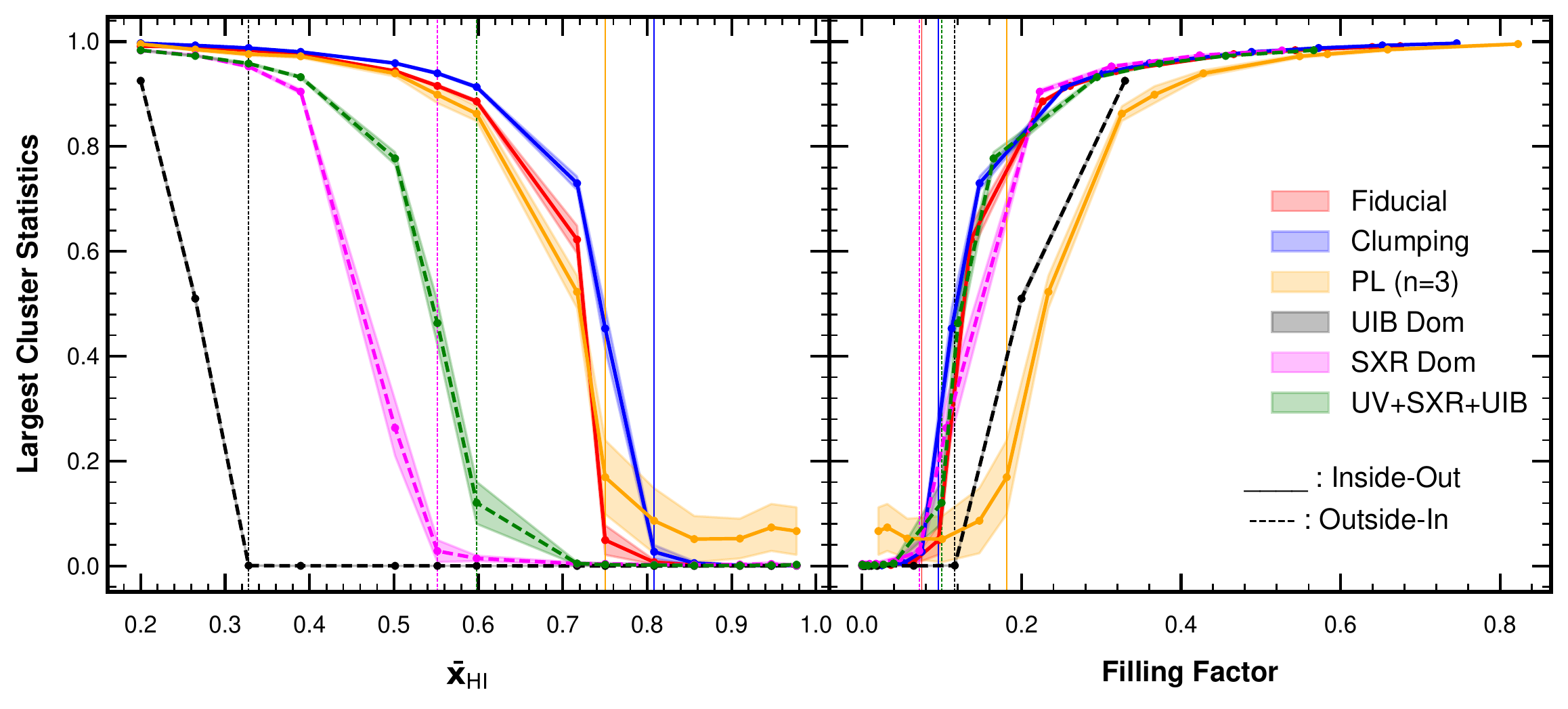}\par 
    \caption{Same as figure \ref{fig:comb_LCS} but with error on LCS, shown by the shaded regions, estimated using the standard deviation of the results from the smaller boxes.}
    \label{fig:LCS_xhi_err}
\end{figure*}

Figure \ref{fig:LCS_xhi_err} shows the LCS as functions of $\xhi$ and filling factor in the left and right panels respectively together with the uncertainties estimated using the standard deviation of the results from the smaller boxes. The solid/dashed curves represent the results for the primary box, and hence same as in figure \ref{fig:comb_LCS}. The uncertainty, shown by the shaded regions with respective colours, is found to be larger before and near the percolation transitions especially for the inside-out scenarios. Particularly for PL ($n=3$) model, we get largest error in the pre-percolation phase since this model produces largest ionized bubbles among all the scenarios tested in this article, hence exhibits more variation in LCS with different realizations.

In summary, we crudely estimate the uncertainty in various quantities studied in the main text by separately analysing the simulation volume divided into 8 equal pieces. As we expected, the results from the primary simulation volume and the smaller ones are statistically consistent with each other. The uncertainties in the quantities like filling factor, number of ionized region per unit volume and most importantly in LCS is larger for inside-out scenarios where larger ionized regions typically forms as compared to the outside-in scenarios. Nevertheless, the estimated errors in LCS for all the scenarios are small enough so that one can easily distinguish the source models; especially between the inside-out and outside-in scenarios.

\section{Shapes of ionized regions in the clumping model}
\label{app:clumping}

\begin{figure*}
    \centering
    \includegraphics[width=0.475\textwidth]{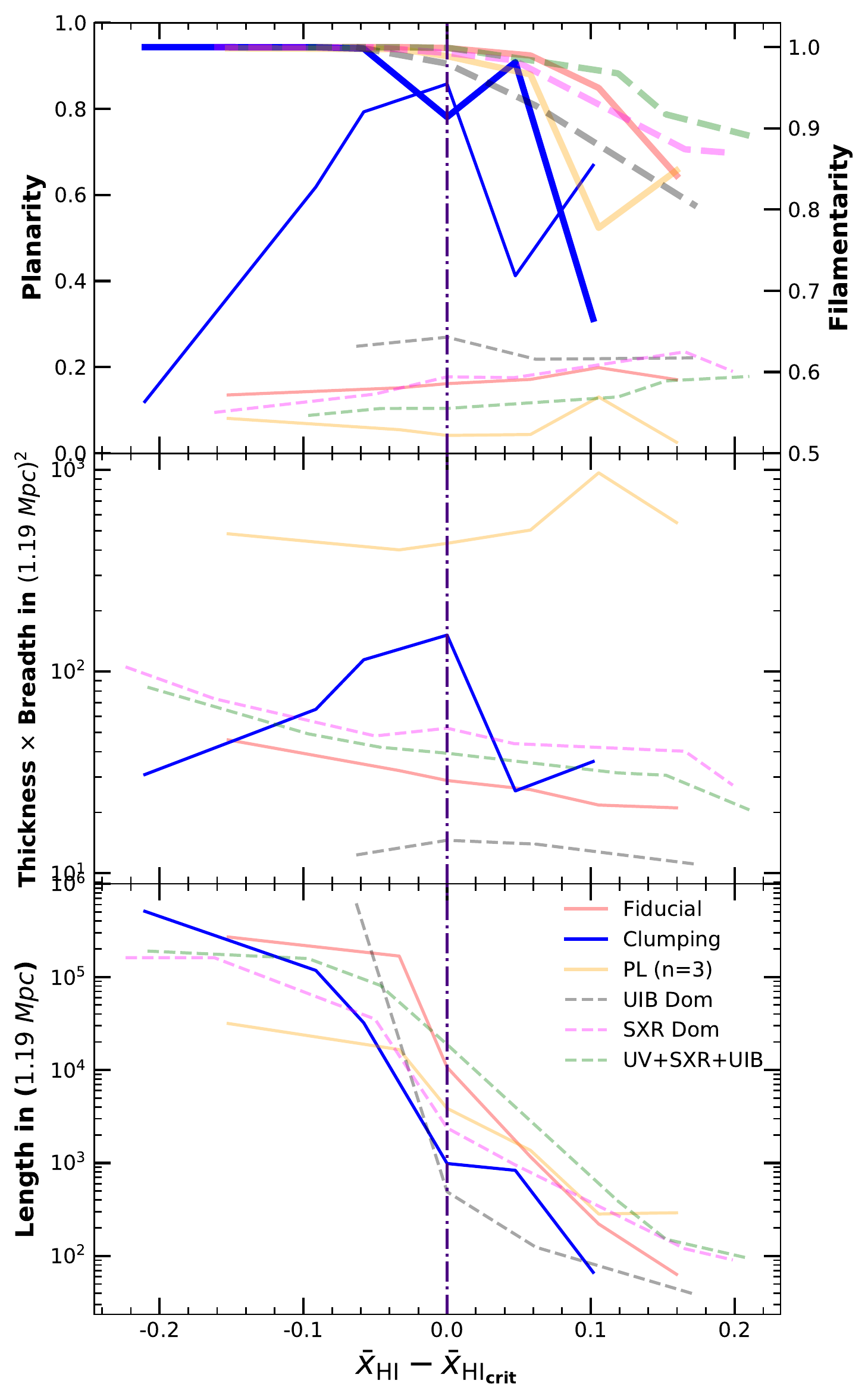}
    \includegraphics[width=0.442\textwidth]{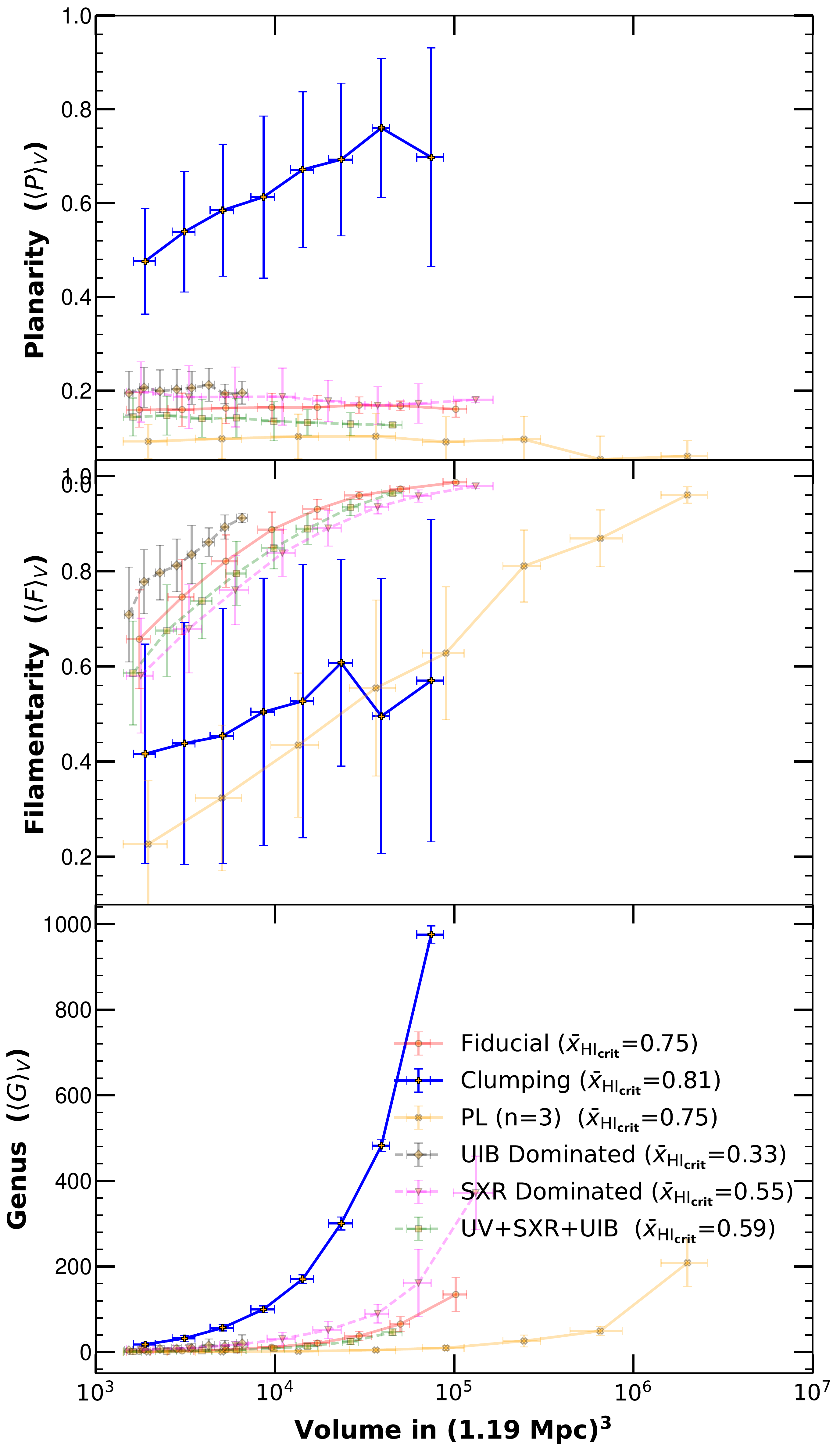}
    \caption{The evolution of the shape of the largest ionized region in the clumping model has been compared with that in the rest of the scenarios in the left panels in terms of planarity, filamentarity (top), cross-section (middle) and length (bottom).
    Note that in the top left panel, we plot planarity (thin curves) along the left y-axis whereas filamentarity (thick curves) is plotted along the right y-axis. The right panels illustrate how the shape distributions of the ionized regions in the clumping model are different from the other scenarios considered in this article.}
    \label{fig:clump_lic_comp}
\end{figure*}

We have seen in section \ref{sec:SF_lir}  that the clumping model in general gives rise to large genus values in the ionized regions because of the fact that the non-uniform recombination produces large number of small neutral pockets that remain neutral till pretty late reionization stage. These neutral pockets form many tunnels increasing the genus values. Another distinctive characteristic of the clumping model is that these tunnels are mostly negatively curved. Thus, the integrated mean curvature (IMC) often becomes very low or even negative sometimes. (In the case of negative IMC, we take its absolute value, see footnote \ref{foot:order_SF}.) This in turn reduces the third Shapefinder in Eq. \eqref{eq:L} (and boosts the second one in Eq. \eqref{eq:B}) often violating the expected natural order: $T\leqslant B \leqslant L$. As explained in footnote \ref{foot:order_SF}, we restore the order by selecting the largest of the three Shapefinders as `length' ($L$) and the smallest one as `thickness' ($T$).
Interestingly, all these translate into higher values of planarity which may not be entirely physical. The boost in planarity of the LIR in the clumping model is more profound at the onset of percolation. However, one can use this behaviour to easily distinguish this type of model with non-uniform IGM from the rest. We demonstrate this in this appendix in more detail.

The evolution of the shape of the largest ionized region (LIR) in the clumping model has been compared with that for the rest of the reionization models in the left panels of figure \ref{fig:clump_lic_comp} (the results for the rest 5 scenarios have been already shown in the main text, in figures \ref{fig:PF_LC} and \ref{fig:TXB_all}). The top-left panel shows the morphological evolution of the LIR in terms of its planarity (plotted by thin curves along the left y-axis) and filamentarity (thick curves plotted along the right y-axis). Like in the other scenarios, the filamentarity of the LIR in the clumping model (thick blue curve) also increases with reionization near percolation. But, the planarity of the LIR is significantly higher in the clumping model (thin blue curve) than that in the rest of the scenarios because of the reasons explained above. Therefore, the morphology of the LIR at the onset of percolation in the clumping model is not strikingly filamentary as opposed to the rest of the models. The middle and bottom panels on the left compare the evolution of the cross-section and the length of the LIR in the clumping model with that in the other scenarios. 
In contrast to other scenarios, the cross-section of the LIR in the clumping model does not remain stable during the percolation transition due to the enhanced planarity. However, the abrupt increase in the length (third Shapefinder) is at par with other scenarios.

The right panels of figure \ref{fig:clump_lic_comp} compare the morphology distribution of the individual ionized regions in the clumping model with that of the rest of the models (these results for the rest 5 scenarios have been already shown in the main text, in figure \ref{fig:PFG}).
The clumping model behaves entirely differently from the rest of the source models in all three panels on the right. Because of the high recombination rate in this model, there exist many pockets of neutral hydrogen which give rise to very high genus values for the ionized regions, as vividly evident from the bottom-right panel. The top and middle panels on the right illustrate the difference in the planarity and filamentarity distributions in the clumping model from the rests. We find overall decreased filamentarity and increased planarity in the ionized regions for the clumping model as compared to the other scenarios. Moreover, we notice large dispersion in the filamentarity and planarity distributions that set the clumping model apart from the rest of the scenarios considered in this work. 
Therefore, in view of the enhanced planarity of the ionized regions in the clumping model with extreme inhomogeneity in the modelling of IGM, one can easily distinguish this type of reionization source mode using the Shapefinders analyses.

\section{Testing with low-resolution fields}
\label{app:low_res}
In the main text, we have analysed a set of simulations with a volume of $714 ~\mpc^3$ and a resolution of $1.19~\mpc$ which is better than that of the data expected from the typical low-frequency surveys in the near future such as the SKA-low. To test how robust our geometrical estimations are, we compare our original results against that from coarse resolution data in this appendix. 
To avoid generating fresh sets of low-resolution 21-cm maps  for different reionization source models, we down-sample our original simulations to reduce the resolution. To be precise, we randomly sample one out of the neighbouring 8 cells and the resulting resolution becomes twice that of the original simulations, i.e. $2.38$ Mpc, keeping the physical size of the boxes the same. This ensures that the low-resolution boxes follow the same reionization history, i.e. the same time evolution of neutral fraction ($\xhi$) and filling factor (FF) for each model, keeping the figure \ref{fig:reion_history} unaltered. Our definition of ionized region ($\rho_{\rm HI}({\bf x})=0$) also remains consistent. 

\begin{figure}
    \centering
    \includegraphics[width=\textwidth]{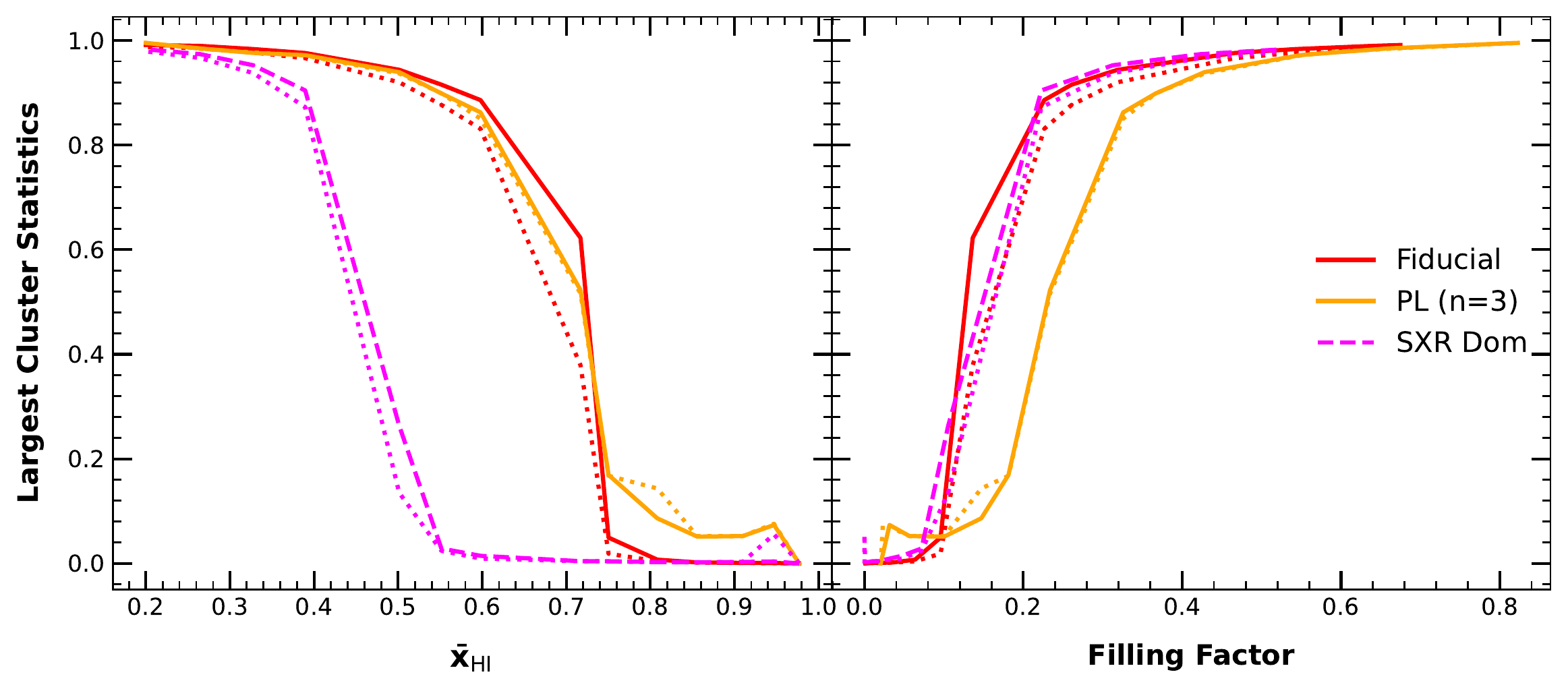}
    \caption{The LCS results from the low-resolution 21-cm maps (dotted curves) have been compared with that from our original resolution maps (solid/dashed curves, same as in figure \ref{fig:comb_LCS}) for three reionization source models. The LCS results are consistent for the two resolutions while the critical $\xhi$ at the onset percolation transitions in all the models remain unchanged.}
    \label{fig:lcs_lowres}
\end{figure}

In figure \ref{fig:lcs_lowres} we compare the evolution of largest cluster statistics (LCS) between the original (solid/dashed curves, same as in figure \ref{fig:comb_LCS}) and low (dotted curves) resolution simulations for three reionization scenarios. The left and right panels show the LCS as functions of $\xhi$ and FF respectively, similar to figures~\ref{fig:comb_LCS} and \ref{fig:LCS_xhi_err}.  We find in both panels that the evolution of LCS for the original and low-resolution simulations are largely consistent in each of the three source models. The critical $\xhi$ at the onset of percolation transition in each model remains unchanged. The slight difference in the LCS curves for the low and high-resolution simulations in the fiducial model arises mostly because we have simulations at sparsely sampled discrete values of redshift or $\xhi$, or in other words, the difference would reduce if we had simulations at finely sampled values of $\xhi$ in the vicinity of the percolation transition. Note that the impact of resolution variation on the LCS evolution is minimum in the PL (n=3) model since the largest ionized region (LIR) in this model is typically the largest among all models with the least genus values (i.e. more simply connected). This ensures that the physical size of the LIR remains mostly unaffected by down-sampling leading to almost unaltered values of LCS.

\begin{figure}
    \centering
    \includegraphics[width=0.5\textwidth]{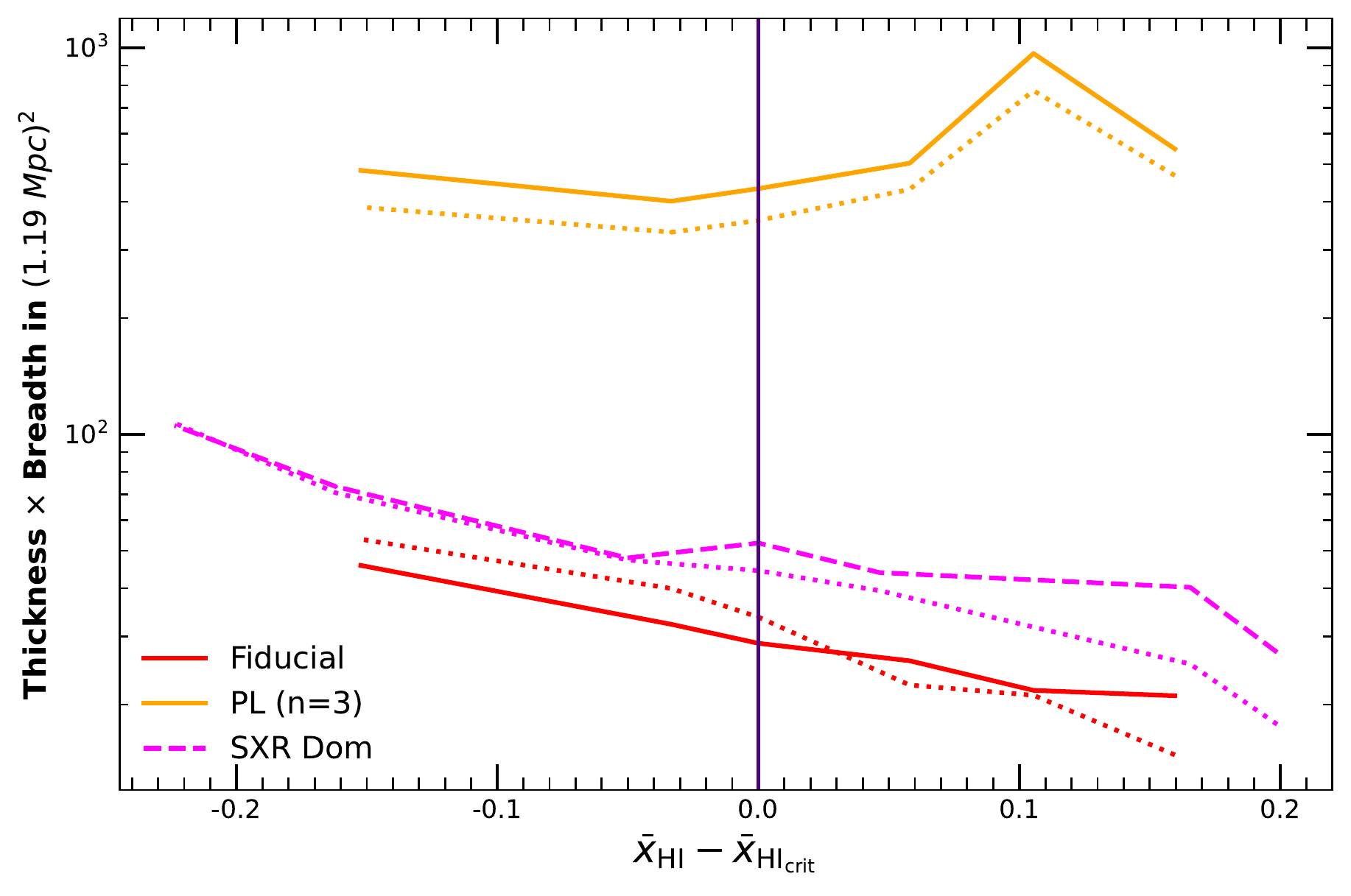}
    \caption{The evolution of the `cross-section' of the largest ionized region have been compared for low-resolution (dotted curves) and the original (solid/dashed curves, same as in the left panel of figure \ref{fig:TXB_all}) resolution fields in three reionization scenarios. Again we find consistent Shapefinder results in the two resolutions.}
    \label{fig:cs_lowres}
\end{figure}

In section \ref{sec:SF_lir}, we show that the cross-section (estimated by $T \times B$) of the largest ionized region (LIR) remains stable during the percolation transition while its `length' increases abruptly. This is found to be valid for all reionization scenarios (except the clumping model) with different characteristic cross-sections which can be used to distinguish the source models. Here, we also investigate the effect of data resolution on this probe.  
We compare the evolution of the cross-section of the LIR for the original and low-resolution simulations in figure \ref{fig:cs_lowres}. The solid/dashed and dotted curves respectively represent the results for original and low-resolution data. Three reionization scenarios are plotted using different colours. We find that the cross-section (of the LIR) estimations are fairly unaffected by reducing the resolution. Thus, we show in this appendix that the ability of LCS and Shapefinder analyses for distinguishing source models is quite robust under reasonable variation in the data resolution.

\bibliographystyle{JHEP}
\bibliography{ref}

\end{document}